\documentclass[fleqn,usenatbib]{mnras}

\usepackage{newtxtext,newtxmath}

\usepackage[T1]{fontenc}
\usepackage{ae,aecompl}

\usepackage{graphicx}	
\usepackage{amsmath}	
\usepackage{amssymb}	
\usepackage[dvipsnames]{xcolor}   

\newcommand\bmB{\boldsymbol{B}}

\newcommand\bmm{{\boldsymbol{m}}}

\renewcommand\bmu{{\boldsymbol{u}}}
\newcommand\bmv{{\boldsymbol{v}}}

\newcommand\imag{\mathrm{Im}}

\newcommand\rmd{\mathrm{d}}

\newcommand\f{\frac}
\newcommand\p{\partial}
\newcommand\cst{\mathrm{constant}}

\makeatletter 
\renewcommand*\env@matrix[1][\arraystretch]{%
  \edef\arraystretch{#1}%
  \hskip -\arraycolsep
  \let\@ifnextchar\new@ifnextchar
  \array{*\c@MaxMatrixCols c}}
\makeatother

\title[Wind-MRI interactions in protoplanetary discs]{Wind-MRI interactions in local models of protoplanetary discs: I. Ohmic resistivity}

\author[Leung \& Ogilvie]{
Philip K. C. Leung$^{1}$\thanks{e-mail: pkcl2@cam.ac.uk}, 
Gordon I. Ogilvie$^{1}$\thanks{e-mail: gio10@cam.ac.uk} 
\\
$^{1}$Department of Applied Mathematics and Theoretical Physics, Centre for Mathematical Sciences, University of Cambridge, \\ 
Wilberforce Road, Cambridge CB3 0WA\\
}

\date{Accepted 2020 July 31. Received 2020 July 27; in original form 2020 July 3}

\pubyear{2020}

\begin{document}
\label{firstpage}
\pagerange{\pageref{firstpage}--\pageref{lastpage}}
\maketitle

\begin{abstract}
A magnetic disc wind is an important mechanism that may be responsible for driving accretion and structure formation in protoplanetary discs. Recent numerical simulations have shown that these winds can take either the traditional `hourglass' symmetry about the mid-plane, or a `slanted' symmetry dominated by a mid-plane toroidal field of a single sign. The formation of this slanted symmetry state has not previously been explained. We use radially local 1D vertical shearing box simulations to assess the importance of large-scale MRI channel modes in influencing the formation and morphologies of these wind solutions. We consider only Ohmic resistivity and explore the effect of different magnetisations, with the mid-plane $\beta$ parameter ranging from $10^5$ to $10^2$. We find that our magnetic winds go through three stages of development: cyclic, transitive and steady, with the steady wind taking a slanted symmetry profile similar to those observed in local and global simulations. We show that the cycles are driven by periodic excitation of the $n=2$ or $3$ MRI channel mode coupled with advective eviction, and that the transition to the steady wind is caused by a much more slowly growing $n=1$ mode altering the wind structure. Saturation is achieved through a combination of advective damping from the strong wind, and suppression of the instability due to a strong toroidal field. A higher disc magnetisation leads to a greater tendency towards, and more rapid settling into the slanted symmetry steady wind, which may have important implications for mass and flux transport processes in protoplanetary discs.
\end{abstract}



\color{black}
\section{Introduction}

Recent observations have shown that protoplanetary discs, believed to be the nurseries of planets, possess a myriad of interesting features \citep{Garufi_etal_2018,Benisty_etal_2015} ranging from concentric rings, to spiral arms, asymmetric features, and more. Many mechanisms have been proposed to explain the dynamics leading to the formation of these structures, including planet-disc interactions \citep{Pinte_etal_2018}, hydrodynamic instabilities \citep{Nelson_eteal_2013}, non-ideal MHD effects \citep{KunzLesur2013,Bai_2014,Suriano_etal_2018} and others \citep{Zhang_etal_2015,Okuzumi_etal_2016,TakahashiInutsuka_2016}. One mechanism, of particular interest recently, is the action of a magnetic disc wind launched because of the presence of a large-scale magnetic field threading the disc \citep{BlandfordPayne_1982}. 
Protoplanetary discs are most likely laminar discs. Their low ionisation profile from their high optical thickness and low temperatures \citep{Gammie_1996} leads to non-ideal MHD effects such as Ohmic diffusion, Hall drift and ambipolar diffusion which suppress the magnetorotational instability (MRI) \citep{BalbusHawley1991a,Hawleyetal1995}, the most commonly invoked mechanism for driving turbulence \citep{Fleming_etal_2000,SanoStone_2002a,WardleSalmeron_2012,Bai_2013,Lesur_etal2014,Bai_2015}. This means that the traditional picture of accretion being driven by turbulent motions acting as an effective viscosity transporting angular momentum radially outwards \citep{ShakuraSunyaev_1973} may not apply to protoplanetary discs. The magnetic disc wind has gained popularity recently as a crucial mechanism for both the accretion \citep{Bai_2016} and ring formation processes \citep{Suriano_etal_2017,RiolsLesur_2019,Riols_etal_2020}.
Such a wind produces large-scale magnetic stresses which can drive significant accretion \citep{Bai_2017,Bethuneetal2017}, while both local \citep{RiolsLesur_2019} and global simulations \citep{Bai_2017,Bethuneetal2017,Riols_etal_2020} of protoplanetary discs with these winds have also exhibited features such as the formation of axisymmetric rings.

\begin{figure}
  \centering
    \includegraphics[width=0.5\textwidth]{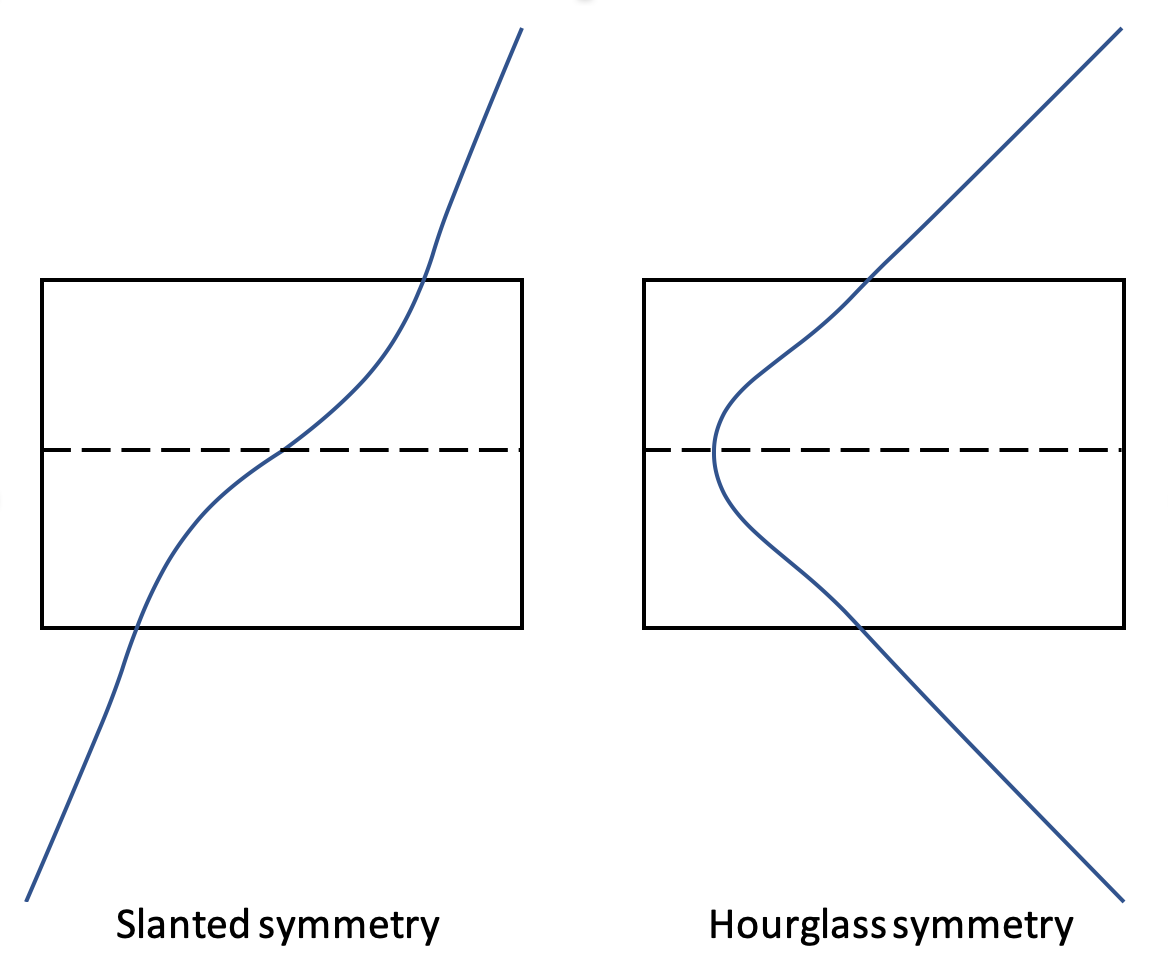}
    \caption{Cartoon illustrating the two disc symmetries found in local simulations. The dotted line denotes the mid-plane, while the blue lines show the shapes of the poloidal magnetic field lines. These have also been found in global simulations.}
    \label{fig:DiscSymmetriesCartoon}
\end{figure}

An interesting aspect of protoplanetary disc magnetic winds brought out by recent simulations is the symmetry of the disc and wind structure about the mid-plane. Traditional models of magnetised discs assumed an `hourglass' symmetry of the magnetic field about the mid-plane (see Figure \ref{fig:DiscSymmetriesCartoon}), where the poloidal field is purely vertical at the mid-plane and bends away from the star above and below. Shearing generates toroidal fields of opposite signs across the mid-plane, while the horizontal velocity fields have the same signs on both sides of the disc, consistent with there being a net accretion flow. However, early local simulations of protoplanetary discs have shown that a `slanted' symmetry state can also develop, where the poloidal field is slanted in one direction at the mid-plane, bending in opposite directions above and below the disc, and a significant toroidal field of a single sign also develops encompassing the whole disc \citep{BaiStone2013,Lesur_etal2014}. This slanted symmetry was later confirmed to occur not only in local simulations (which inherently do not distinguish between the radial inward and outward directions in relation to the star), but also in global simulations. The same features as the local slanted solution were observed in the disc region and extending to the lower atmosphere, before (in some cases) a kink occurs in the upper atmosphere bending the field in the half of the disc that is slanted towards the star back outwards \citep{Bai_2017,Bethuneetal2017,Riols_etal_2020}. In both local and global cases, the properties of the wind and the disc are significantly affected by which symmetry the solution takes. In the local scheme, a slanted symmetry solution implies no net accretion or magnetic flux transport because of a cancellation of the contributions from the upper and lower halves of the disc. For global solutions, even in cases where the field eventually bends back outward from the star in the upper atmosphere, both the disc wind and accretion flow are significantly changed (becoming highly one-sided), and the overall radial transport of vertical flux is also greatly affected \citep{Bai_2017,Bethuneetal2017}.

Thus far, there has been no agreed explanation for the development of the slanted symmetry state. Previous authors \citep{Bai_2017,Bethuneetal2017} have invoked non-ideal MHD effects such as the Hall-shear instability \citep{Kunz_2008}, which arises from the presence of Hall drift, to explain the development of the strong radial and toroidal net flux characteristic of the slanted solution. Others \citep{Gressel_etal_2020} have suggested that it could be a manifestation of corrugation of the mid-plane by the vertical shear instability (VSI) \citep{Urpin_2003,UrpinBrandenburg_1998}. However, settling to the slanted symmetry state was also observed in global simulations where Ohmic and ambipolar diffusion were the only non-ideal effects present \citep{Gressel_etal_2020}, and even in purely Ohmic \citep{Rodenkirch_etal_2020} or ambipolar \citep{Riols_etal_2020} discs. The latter simulation is also locally isothermal, shutting down the VSI and ruling it out as necessary for the development of the slanted symmetry. 

A possible mechanism suggested which could affect the disc wind configuration is the role of the MRI in the wind generation process \citep{SuzukiInutsuka_2009,Suzuki_etal_2010,Ogilvie2012}. The MRI is a shear-induced instability which converts rotational energy in the disc to magnetic energy by amplifying the horizontal fields \citep{BalbusHawley_1998}. \citet{Lesur_etal_2013} first did a detailed investigation of the link between the MRI and magnetic winds using local simulations, and showed that large-scale channel modes could naturally produce steady outflows in the nonlinear regime. A further investigation by \citet{Riols_etal_2016} uncovered how these modes can also drive wind cycles with periodic outbursts in discs near the MRI marginal stability boundary. Since MRI channel modes naturally take either a slanted or hourglass symmetry about the mid-plane \citep{SanoMiyama_1999,Latter_etal_2010}, they may also have a link to the wind configurations that we see in protoplanetary disc simulations. Past work on wind-MRI interactions has so far focused on the ideal MHD regime \citep{Lesur_etal_2013} and much stronger fields \citep{Riols_etal_2016} than are usually considered in protoplanetary discs \citep{Wardle_2007,GuiletOgilvie2014}. In this paper, we aim to see if such interactions are also relevant in the non-ideal MHD weak field regime more suitable for modelling protoplanetary discs. We would like to find out to what extent large-scale MRI dynamics may influence the launching and configuration of magnetic disc winds, and how the wind in turn feeds back on the development of the MRI. Our ambition ultimately is to characterise and predict what disc symmetries and wind solutions may develop based on the different MRI modes being excited, and to provide a greater understanding into the development of the slanted symmetry steady wind seen in global simulations. We would also like to examine how factors such as disc magnetisation may affect the evolution and outcome of disc configurations.

For that purpose, we performed radially local 1D vertical shearing box simulations of stratified discs with a net vertical magnetic field using the PLUTO code. Although not all physical processess (such as turbulence) are present because of its 1D and local nature, the model is nevertheless sufficient to capture the large-scale channel modes we are after. Another advantage of a local model is that we can explore a much wider parameter space and run simulations for much longer than global ones, allowing us to access the long-term outcomes of the wind. As slanted symmetry profiles have been reported even in simulations with only Ohmic resistivity \citep{Rodenkirch_etal_2020}, we restrict the non-ideal physics we consider to Ohmic resistivity only, to identify the minimum ingredients required for the disc to adopt the different wind symmetries. While Hall drift and ambipolar diffusion also have significant impact on protoplanetary disc dynamics \citep{Lesur_etal2014,Bai_2015}, their anisotropic and nonlinear nature makes it much harder to isolate and evaluate their effects, and we relegate their study to a future work. On the other hand, an Ohmic only regime may also be appropriate for the inner disc regions where it is significantly stronger than Hall drift and ambipolar diffusion \citep{Wardle_2007,Bai2011a}. Finally, we used targeted triggering of MRI modes through our initial conditions to better understand their effects on the wind topology. Using this approach, we examined whether the history of the disc is important to the intermediate and long term outcomes of the wind solution, and assessed the varying importance of MRI modes of different morphologies.

This paper is organised as follows: In Section \ref{sec:Model}, we describe the model and justify our use of the 1D local scheme. In Section \ref{sec:setup}, we explain our setup, particularly the tall boxes we used as well as our modelling of the non-ideal physics. Section \ref{sec:wind_solutions} details the initial conditions we used for the targeted triggering of MRI modes, the different categories of wind solutions we found: cyclic, transitive and steady, as well as a brief description of their evolution and their dependence on disc parameters. In Sections \ref{sec:cycles-interpretation}, \ref{sec:trans} and \ref{sec:bulge-sat}, we investigate and propose the mechanisms behind the cyclic and transitive states and saturation to the steady state wind respectively. We summarise our results in Section \ref{sec:discussion}, and discuss how they relate to wind solutions found in other simulations and possible astrophysical applications.

\section{Model and equations}
\label{sec:Model}

We use the standard Cartesian local shearing-sheet description \citep{Goldreich&LyndenBell_1965} to investigate the behaviour of a radially local patch of the disc. The $x$, $y$, and $z$ coordinates represent the radial, azimuthal and vertical directions respectively. We assume that variables do not vary in the horizontal directions ($\p / \p x , \p / \p y = 0$), motivated by the laminar vertical 1D profiles found in both local \citep{Bai_2013} and global simulations \citep{BaiStone2017,Bai_2017}. This assumption inevitably reduces the complex physics that may occur in a real disc, but may be sufficient to capture the essential mechanisms that influence wind launching.

For simplicity, we assume an isothermal disc with equation of state $p = c_\text{s}^2 \rho  $, where $c_\text{s}$ is the sound speed and is uniform in the domain. The system of equations governing the development of the density $\rho$, velocity $\bmv$ and magnetic field $\bmB$ under these approximations is then \citep{Ogilvie2012}
\begin{equation}
    \f{\p \rho}{\p t} + \f{\p ( \rho v_z)}{\p z} 
    = \varsigma (z,t),
\end{equation}
\begin{equation}
    \rho \left( \f{\p v_x}{\p t} + v_z \f{\p v_x}{\p z} - 2 \Omega v_y \right)
    = \f{\p}{\p z} \left( \f{B_x B_z}{\mu_0} + \rho \nu \f{\p v_x}{\p z} \right),
\end{equation}
\begin{equation}
    \rho \left(\f{\p v_y}{\p t} + v_z \f{\p v_y}{\p z} + \f{1}{2}\Omega v_x \right)
    = \f{\p}{\p z} \left( \f{B_y B_z}{\mu_0} + \rho \nu \f{\p v_y}{\p z}\right),
\end{equation}
\begin{equation}
    \rho \left( \f{\p v_z}{\p t} + v_z \f{\p v_z}{\p z} \right)
    = \rho g_z - \f{\p}{\p z} \left( c_\text{s}^2 \rho + \f{B_x^2 + B_y^2}{2 \mu_0} 
    - \f{4}{3} \rho \nu \f{\p v_z}{\p z} \right),
\end{equation}
\begin{equation}
    \f{\p B_x}{\p t} = 
    \f{\p}{\p z}
    \left( v_x B_z - v_z B_x \right)
    + \f{\p}{\p z} \left( \eta \f{\p B_x}{\p z} \right),
\end{equation}
\begin{equation}
    \f{\p B_y}{\p t} = - \f{3}{2} \Omega B_x
    + \f{\p}{\p z} \left( v_y B_z - v_z B_y \right)
    + \f{\p}{\p z} \left( \eta \f{\p B_y}{\p z} \right),
\end{equation}
where $\eta$ is the Ohmic diffusivity, and is allowed to vary with height, and $\nu$ is the kinematic viscosity. $B_z$ is a constant parameter of the 1D model, because of flux conservation. We can define $H=c_s/\Omega$ as the standard hydrostatic scale-height of the disc, while our unit of time is given by $\Omega^{-1}$. The source term $\varsigma(z,t)$ in the continuity equation represents an artificial mass injection that replenishes mass lost to the wind. In a real disc, this mass would be replenished by radial flows from neighbouring parts of the disc, but these depend on radial gradients and curvature effects that are not represented in the shearing sheet model. The various mass replenishment schemes used in this paper are outlined in Section \ref{sec:mass_repl_scheme} and their effects on our results discussed in Section \ref{sec:mass_repl_effects}. 

As noted in \citet{Riols_etal_2016}, the shearing box gravity term $g_z = -\Omega^2 z$ is only appropriate when the vertical scales of interest are small compared to the disc radius. However, when $z \gg H$, this term actually completely suppresses vertical outflow from the box, as the gravitational potential well becomes infinite. We modify the vertical gravity in the same manner as \cite{Riols_etal_2016}, taking into account the finite distance from the central object such that
\begin{equation}
    g_z = - \f{GMz}{(r_0^2+z^2)^{3/2}}
    = - \Omega^2 H \f{\Hat{z}}{(1+\delta^2\Hat{z}^2)^{3/2}},
\end{equation}
where $\Hat{z}\equiv z/H$ and $\delta = H/r_0$, with $r_0$ being the radial location from the star. Note that $\delta=0$ brings us back to the standard shearing box gravity term, leading to the Gaussian hydrostatic density profile. When $0<\delta<1$, hydrostatic equilibrium is obtained by integrating with respect to $z$ the $z$ momentum equation,
\begin{equation}
    \f{\rmd \rho}{\rmd \Hat{z}} 
    = - \f{\rho \Hat{z}}{(1+\delta^2 \Hat{z}^2)^{3/2}} , 
\end{equation}
giving us a modified solution
\begin{equation}
    \rho(z) = \rho_0 \exp{\left[ - \f{1}{\delta^2}
    \left( 1 - \f{1}{\sqrt{1 + \delta^2 \Hat{z}^2}}\right)\right]},
\end{equation}
where $\rho_0$ is the density at the mid-plane.
Using binomial expansion, this solution can be shown to tend towards the standard Gaussian hydrostatic equilibrium when $\delta \Hat{z} \ll 1$, in other words when $z \ll r_0$. When $z\to\infty$, $\rho(\infty)$ differs from the Gaussian solution by settling at a floor value of $\exp{(-1/\delta^2)}$ instead of vanishing to $0$.

Similar modified gravity terms to ours have been used by other authors in both modelling accretion \citep{Matsuzaki_etal_1997} and galactic discs \citep{KuijkenGilmore_1989}. Although a full treatment should in theory also account for the variation of the radial gravity term at large scale heights, as a simplification we assume that this is not important for the flow dynamics we are studying, and only apply the vertical gravity modifications in our 1D models. Recent observations of T-Tauri stars suggest that the typical $\delta$ for protoplanetary discs is between $0.03$ and $0.2$ \citep{Andrews_etal_2009,Grafe_etal_2013}. Unless otherwise stated, we chose $\delta=0.033$ in our simulations to represent a typical protoplanetary disc. 

\section{Numerical setup and parameters}
\label{sec:setup}

\subsection{Numerical code}

We used the shearing box module of the astrophysical code PLUTO, developed by \cite{Mignone_etal_2007}. The compressible MHD equations are integrated in their conservative form using a finite-volume method with a Godunov scheme. The fluxes are computed by the HLLD Riemann solver unless otherwise stated, and we found no significant differences to our results when we varied the solver. Time stepping is done using a Runge-Kutta method of third order. 

\subsection{Boundary conditions and mass replenishment}

Simulations are done for the whole vertical extent of the disc, with both sides of the disc mid-plane explicitly calculated. This is distinct from the approach of \citet{Riols_etal_2016}, where symmetry was imposed with respect to the disc mid-plane and simulations were restricted to the upper half of the disc. Our approach allows us to explore geometries different from the classical hourglass symmetry wind-launch configuration, as discussed in the introduction. Local simulations \citep{BaiStone2013,BaiStone2014,Bai_2015} found that discs may settle into the slanted symmetry, whether in the ideal MHD regime or not. Although such a configuration would not be physical at large $z$, as it would imply that one part of the field is bending towards the star, recent global simulations \citep{BaiStone2017,Bai_2017,Bethuneetal2017,Gressel_etal_2020} have suggested that in certain radial locations such a symmetry is indeed adopted throughout the vertical extent of the disc region before field lines bend back in the normal manner away from the star further up in the atmosphere. Hence it would be useful to relax the symmetry assumptions of the solution to explore what factors contribute to the disc adopting a particular configuration.

Following \cite{BaiStone2013}, we use an outflow boundary condition in the vertical directions that has zero vertical gradient for velocity and magnetic fields, while density is attenuated following the Gaussian profile to account for vertical gravity. This attenuation significantly reduces the excitation of spurious artificial waves near the boundary. \citet{Lesur_etal_2013} noted that care is needed in implementing the boundary conditions for the magnetic field, as they found that using a zero vertical gradient condition prevented an outflow from being launched. However, this was not found to be the case in our simulations. An explanation for this may be that all the outflows in our simulations are super-Alfv\'enic, and therefore much less sensitive to the field configuration at the boundary than some of the sub-Alfv\'enic outflows they were investigating. Nevertheless, we ran simulations using both a vertical field boundary condition, and fixing the horizontal fields to finite values at the boundary, and found them to have negligible impact on our results.

\label{sec:mass_repl_scheme} In a global disc, radial redistribution of the material would replenish mass in a local patch that is lost to the wind. We mimic this in our local model by injecting mass near the mid-plane at each numerical time step. In the system of equations, this is equivalent to adding a source term in the mass conservation. We use the same source term as prescribed by \cite{Lesur_etal_2013},
\begin{equation}\label{eq:Lesur_mass_src}
    \zeta(z,t) = \f{2 \Dot{m}_i(t)}{\sqrt{2\upi}z_i}
    \exp{\left( - \f{z^2}{2 z_i^2} \right)},
\end{equation}
where $\Dot{m}_i(t)$ is the mass injection rate, and $z_i$ controls the width of the region about the mid-plane where most of the mass replenishment occurs. For most simulations, we replenish the mass such that a constant disc surface density $\Sigma = \int^{L_z}_{-L_z} \rho \rmd z$ is maintained in time, but we also explored the effect of other schemes, described in Section \ref{sec:mass_repl_effects}. It is important to note though that such injection of mass breaks momentum conversation in the shearing box, as the mass is injected with the local velocity (injecting momentum so that the velocity stays the same), while it leaves the domain with a different velocity at the upper and lower boundaries. The loss of horizontal momentum from the box (including that from a torque exerted at the vertical boundaries by the Maxwell stress), drives a mean horizontal flow, which was interpreted by \cite{Riols_etal_2016} as the accretion flow for the $x$ component, together with a small departure from Keplerian motion for the $y$ component.

\subsection{Box size and resolution:}

As pointed out by previous authors \citep{Fromangetal2013,Lesur_etal_2013}, the choice of box size (which we label here as $2 L_z$, with $L_z$ being the maximum height above the mid-plane) and boundary conditions can have a strong effect on the wind solution obtained. This is especially the case when critical points of the wind flow (see definitions in \citet{Ogilvie2012}) lie outside the simulation domain, allowing information to be propagated from the box boundaries back to the disc and affecting its behaviour. For the weak field strengths we consider in our simulations, the slow magnetosonic point and Alfv\'en point are always crossed within the box as long as $L_z \gg H$. However, like the simulations of \citet{Riols_etal_2016} and in line with other studies \citep{Lesur_etal_2013}, we are unable to find solutions that pass through the fast magnetosonic point. Hence it is possible that the vertical boundaries still have an effect on the nature of our wind solutions, although we find that the properties of our steady wind solutions converge with increasing box height, while the phenomenology of the wind behaviours is also independent of box size as long as $L_z\gg H$ (see Section \ref{sec:box_size_effects}).

Another cause of non-convergent wind properties with increasing box size in traditional shearing boxes is the nature of the standard shearing box gravity term being linear in $z$, leading to the gas being trapped in an infinite potential well. This effect has largely been mitigated through the use of the modified gravity term we have adopted from \citet{Riols_etal_2016}, and becomes negligible when $L_z/H > 1/\delta $, which in our case of $\delta = 0.033$ translates to $L_z > 30 H$.

We mainly use two box sizes for our simulations. The first is a relatively `small' box of $L_z=12 H$, while the second is a `large' box with $L_z=70 H$. The latter is chosen as it satisfied the considerations outlined above with the exception of crossing the fast magnetosonic point for the parameter space we explore. However, running simulations in such tall boxes is costly, as they require a large number of grid points to resolve the dynamical features appropriately. We find that the phenomenological behaviour of the wind states in the tall box are the same even in much smaller boxes. Since our interest in this paper is in gaining an understanding into the mechanisms behind the generation of these wind states, rather than trying to predict the precise properties of real discs, we use the `small' box simulations to further our exploration of the parameter space and their effects on disc behaviour.

For the $L_z=12H$ small boxes, we use 200 grid points to resolve the mid-plane region $\lvert z \rvert < 2 H$, while the two atmospheric regions $\lvert z \rvert > 2H$ are spanned by 500 points each. For large boxes with $L_z = 70 H$, we also use 200 grid points to resolve the mid-plane region $\lvert z \rvert < 2 H$, while the atmospheric regions $\lvert z \rvert > 2H$ have 2400 grid points each. The finer grid in the mid-plane region is motivated by small-scale structures like acoustic waves that arise more naturally near the mid-plane. We vary the resolution to make sure that solutions are not drastically affected by the values we have chosen.

\subsection{Physical parameters}

The surface density in all simulations is fixed to be equivalent to that of a hydrostatic disc with mid-plane density $\rho_0=1$, which sets our unit of mass. We use units such that $\mu_0,c_s$ and $\Omega$ are set to $1$. The magnetic field $B_z$, independent of $z$ and $t$ in the shearing box formulation, is derived from the mid-plane $\beta_z$ (ratio of gas pressure to magnetic pressure),
\begin{equation}
    \beta_{z0} \equiv \f{2 \mu_0 c_\text{s}^2 \rho_0 }{B_z^2},
\end{equation}
which we set as a dimensionless parameter for the problem. 

As we are primarily interested in investigating how the general shape of the diffusivity profile affects the phenomenology of the disc and wind, it is not necessary for us to solve the complex chemical networks to determine a precise profile for the resistivity. In a protoplanetary disc the resistivity is high near the mid-plane but much lower in the atmosphere, where the signficant ionisation due to FUV radiation and X-ray heating lead to near-ideal MHD conditions. To mimic this situation, we use a simplified analytic $\eta$ profile which has a fixed constant value in the disc mid-plane region, before decaying exponentially to a floor value in the atmosphere. Mathematically, this is given by
\begin{equation}
    \eta = 
    \begin{cases}
    ~~ \eta_{0},
    & \lvert z \rvert < z_{\text{c}}, \\
    ~ (\eta_{0}-\eta_\infty )
    \exp{[-5(\lvert z \rvert - z_{\text{c}} )]}
    + \eta_{\infty},
    & \lvert z \rvert > z_{\text{c}} .
    \end{cases}
\end{equation}
$\eta_0$ and $\eta_\infty$ are the mid-plane and atmospheric diffusivity values respectively, while $z_\text{c}$ sets the height at which the transition occurs. To estimate suitable values to use for our simulations, we used the ionisation model of \citet{Lesur_etal2014} coupled with accounting for dust-enhanced recombination from \citet{BethuneLatter_2020} (where the ionisation fraction further lowered by a factor of $10^2$) to yield mid-plane $\eta_0$ values of $5.8$ $H^2 \Omega$ and $1.07 \cdot 10^{-2}$ $H^2 \Omega$ at disc radii $R = 1$ and $5$ AU respectively. Ionisation calculations (in the absence of FUV radiation) indicate an increase in the ionisation fraction, $x_e$, by a factor of between $10^{2}$ and $10^{4}$ from the mid-plane to the atmosphere at these radii (see Figure 1 of \citet{BethuneLatter_2020}). This in turn corresponds to a decrease of between $10^{-2}$ and $10^{-4}$ in the resistivity, which varies as $x_e^{-1}$. When FUV is included, this increases the ionisation fraction in the upper regions beyond $\lvert z\lvert = 4H$ even further to near ideal MHD conditions. For most simulations, we used $\eta_0 = 2 ~ H^2 \Omega$, representative of the conditions in the inner disc, while we varied the floor value $\eta_\infty$ from $1/200$ that of the mid-plane value (for most simulations), down to $0$ to examine the effects of ideal MHD atmospheric conditions on the solutions we obtain. $z_c$ is set to $2H$ for all simulations, which follows the ionisation depth for FUV photons estimated in \citet{Simon_etal2015} for the $\eta_\infty=0$ case, while this cut-off height also corresponds well with the ionisation profile of \citet{BethuneLatter_2020} in the absence of FUV when the higher $\eta_\infty$ value of $0.005 ~ H^2 \Omega$ is used. 

We mostly assume inviscid discs, motivated again by the laminar protoplanetary disc solutions recovered in local and global simulations. For simulations with large boxes ($L_z > 30 H$), we found that strong numerical instabilities appear near the upper and lower boundaries in our steady wind solutions. In order to avoid these instabilities, we followed the prescipriton of \citet{Riols_etal_2016} by introducing a small, uniform dynamic viscosity $\rho \nu$, such that the kinematic viscosity $\nu \propto 1/\rho$ has the value $10^{-5}$ in the mid-plane but increases with $\lvert z \rvert$. We found that while the general shape of the solution is not changed significantly by this addition, numerical instabilities are indeed smoothed out when the viscosity is included. As noted in \citet{Riols_etal_2016}, this prescription might also be physically relevant in the isothermal case, as $\nu$ can be estimated as the product of the thermal velocity and the mean free path, which scales as $1/\rho$ \citep{Maxwell_1866}. For our runs with $L_z < 30 H$ we used a zero viscosity treatment, while we only added the artificial viscosity for runs with $L_z>30 H$ as the solutions tend towards the steady state.

\section{Categorisation of wind solutions}
\label{sec:wind_solutions}

\subsection{Initial conditions and obtaining a solution}
\label{sec:InitialConditions}

\begin{figure}
  \centering
    \includegraphics[width=0.5\textwidth]{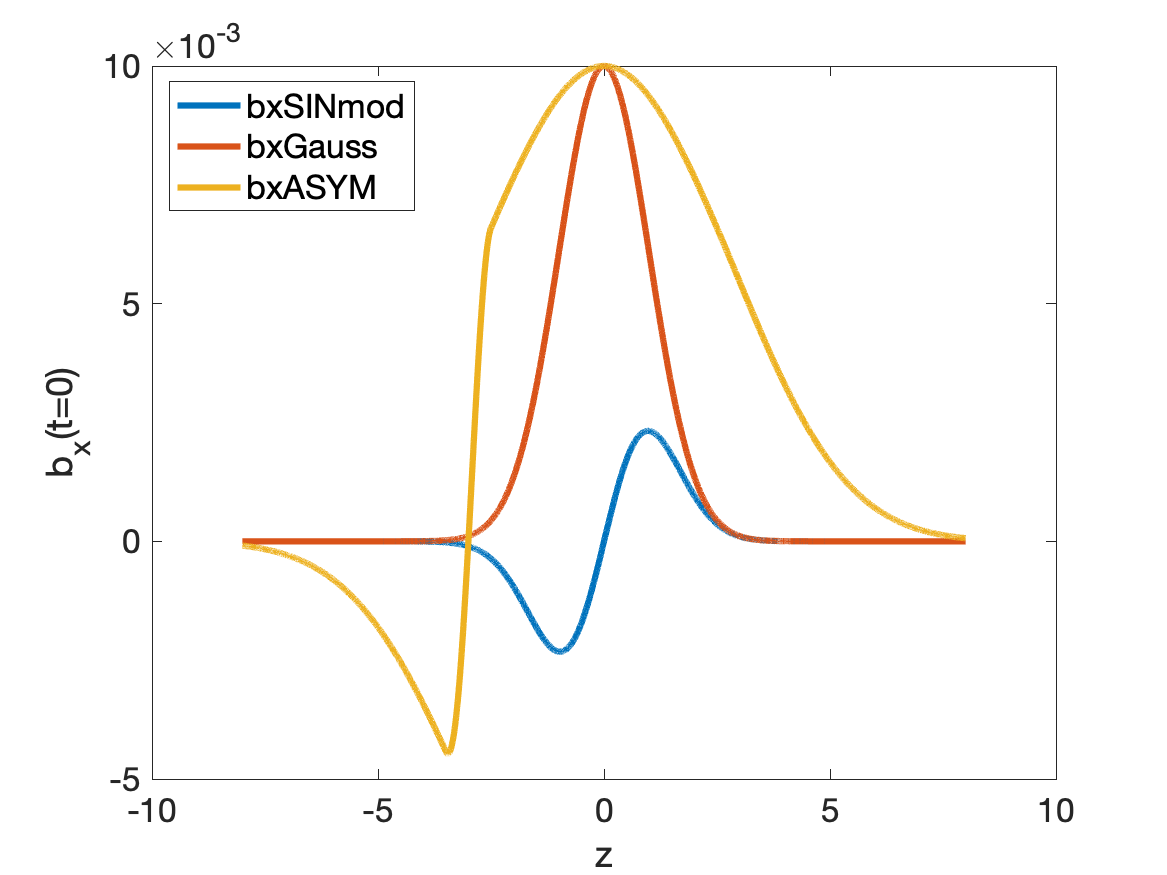}
    \caption{Initial $b_x$ profiles, see Section \ref{sec:InitialConditions} for their analytic forms.}
    \label{fig:bxINITs}
\end{figure}

It is not practical to initialise a simulation in a tall box with a small $\delta$ value from a hydrostatic equilibrium state, because the very low density in the atmosphere leads to a very high Alfv\'en speed that forces the time-step to be extremely small. We used two different methods to obtain wind solutions (which have much higher atmospheric densities than the hydrostatic state) in our extended boxes. The first loosely follows the prescription of \citet{Riols_etal_2016}. We start with a medium box of size $L_z=15$ and a high $\delta$ value of 0.33, which leads to a floor density of $1.028\cdot 10^{-4}$ in the modified hydrostatic equilibrium. The disc is embedded in a vertical field of a strength corresponding to the value of $\beta$ that we wish to investigate. Small random perturbations in the velocity profiles are then introduced, which are amplified by the MRI instability. As the wind solution develops, the density profile becomes more spread out, reaching above $ 10^{-3}$ at the boundaries. After the solution has reached steady state, we slowly reduce $\delta$ back to our desired value of $0.033$, while we extend the size of the box gradually by uniform extrapolation of the boundary values, and allowing each model to settle into the new equilibrium wind solution. Using this method, we were able to obtain the slanted symmetry steady states described in Section \ref{sec:unphysical_steady}. However, unlike in \citet{Riols_etal_2016}, we did not find that the wind solution bifurcates to a periodic outflow as the box size was increased beyond a certain height. Rather, solutions initiated with steady state profiles of smaller boxes always relaxed to the same type of steady state profiles, with converging wind properties.

The second method began with a medium box of size $L_z = 12$ and the desired $\delta$ value of $0.033$. Again, we used a hydrostatic disc threaded by a vertical field as the initial condition, except this time we arbitrarily added a small fraction of the mid-plane density to the entire disc, to avoid the high Alfv\'en speeds that lead to impractically small time-steps in the atmosphere. For most simulations, a value of $10^{-4}$ in code units was chosen, motivated by the typical density measured by \citet{Riols_etal_2016} at their upper disc boundaries.

Instead of using random velocity perturbations, we started the simulations with three different profiles of $B_x$ to examine the excitation of MRI modes of different symmetries, and its effect on the wind solutions obtained. The first, denoted `bxSINmod', has the form
\begin{equation}
    B_x (t=0) = 0.01 B_z \sin{[k_z z]} \exp{(-z^2/2)} ,
\end{equation}
where $k_z = 2 \upi / L_z$ is the wavenumber of a complete wave across the vertical domain, 
giving the initial profile an hourglass symmetry about the mid-plane. The second profile, `bxGauss', explores perturbations with a slanted symmetry, and is simply a Gaussian function,
\begin{equation}
    B_x (t=0) = 0.01 B_z \exp{(-z^2/2)}.
\end{equation}
The third, denoted `bxASYM', explores the effect of starting with an asymmetric profile about the mid-plane but with an hourglass geometry in the atmosphere, motivated by the asymmetric steady state profiles observed in both local and global simulations \citep{BaiStone2013,Bai_2017}. It has the form
\begin{equation}
    B_x (t=0) = 0.01 f(z) B_z  \exp{(-z^2/15)},
\end{equation}
where
\begin{equation} 
    f(z) = 
    \begin{cases}
        - \cos(k_z(z+3.5)/2), & z < -3.5 \\
        \cos(\upi(z+2.5)), & -3.5 < z < -2.5 \\
        1 , & \lvert z \rvert < 2.5 \\
        \cos(k_z(z-2.5)/2), & z > 2.5
    \end{cases}.
\end{equation}
A plot showing these initial $B_x$ profiles is shown in Figure \ref{fig:bxINITs} (note the lower case `b' denotes normalisation with respect to $B_z$). 

Under these conditions, we found that solutions relax, depending on the initial conditions used, to one of two cyclical states that persist for $100$s of orbits, before a growing mid-plane perturbation slowly transited the disc to the steady state profile of slanted symmetry obtained using the first method. We restarted simulations from both the cyclic states and the slanted symmetry steady state in taller boxes with constant extrapolation to examine the effect of the extended vertical domain. We found that the same type of wind behaviour is retained, with the solution converging to an extended version of either the cyclic or slanted symmetry steady wind solutions of the smaller box runs. For the cyclic solutions, an eventual convergence to the steady wind solution is then again observed after a timescale of $100$s of $\Omega^{-1}$.

{\color{black}Note that the runs are named such that the numbers after `b' denote the $\beta_0$ value, while the letter after the underscore denotes the initial $b_x$ profile, with `A' for `bxASYM', `S' for `bxSINmod' and `G' for `bxGauss'. Unless otherwise labelled in the name, all runs have $\delta=0.033$, $\Hat{z}_i=0.5$ and $L_z=12H$. For example, \textbf{b1e5\_S} has $\beta_0=10^5$ and is initiated from the `bxSINmod' profile, while \textbf{b200\_G} has $\beta_0=200$  and is initiated from the `bxGauss' profile.}

\subsection{Phenomenology of wind solutions}

\begin{figure*}
  \centering
    \includegraphics[width=\textwidth]{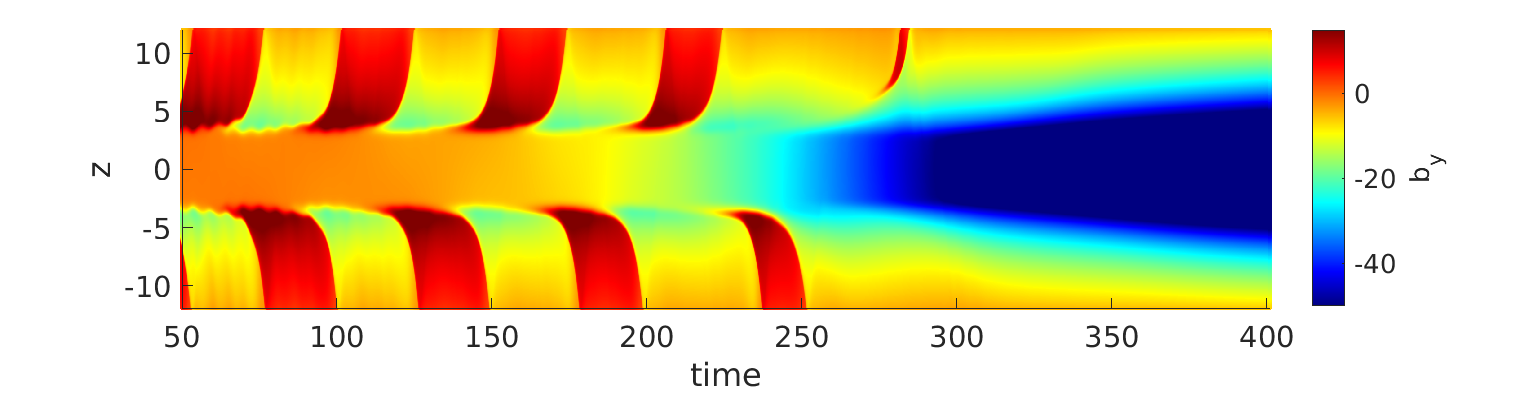}
    \includegraphics[width=\textwidth]{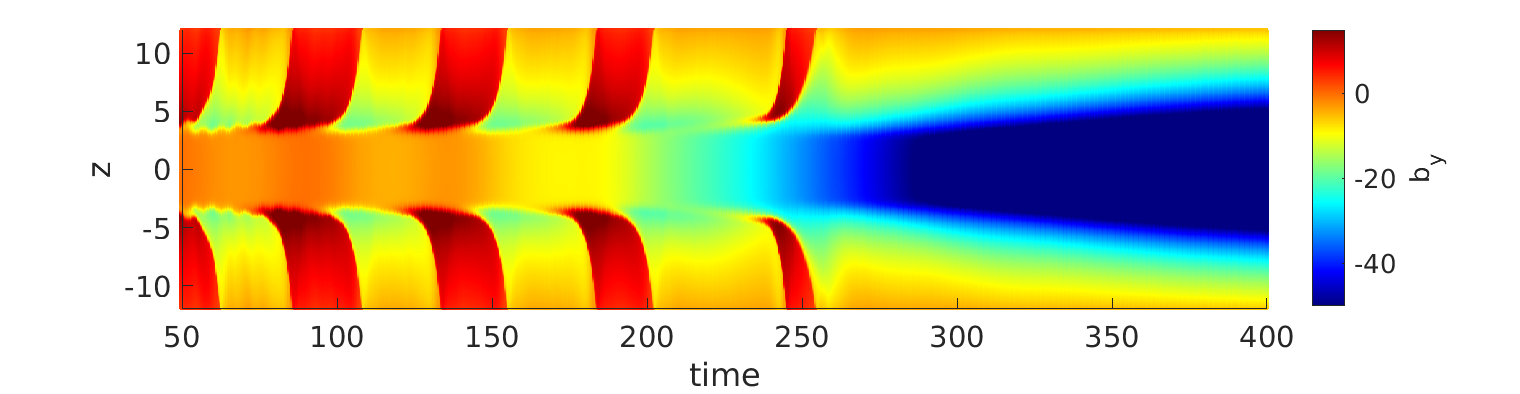}
    \caption{Space-time plots of $b_y$ for the situations where: (top) an hourglass symmetry cyclic state transits to the slanted symmetry steady state, and (bottom) a slanted symmetry cyclic state transits to the slanted symmetry steady state. The first $50$ time units are not displayed as the runs are still in a chaotic transient phase with drastic variations that are strongly dependent on the particular initial conditions used. The runs depicted are \textbf{b1e5\_A} and \textbf{b1e5\_G} respectively.}
    \label{fig:b_y-spacetime-general}
\end{figure*}

In our simulations, the wind solutions obtained can be put into four general types: (i) a cyclic solution with hourglass (odd-$z$ in $b_{x,y}$) symmetry about the mid-plane, (ii) a cyclic solution with slanted (even-$z$ in $b_{x,y}$) symmetry about the mid-plane, (iii) a cyclic to steady wind transition state, and (iv) a steady wind solution with slanted (even-$z$ in $b_{x,y}$) symmetry about the mid-plane. All four types of behaviour can be seen in the space-time plots of $b_y$ in Fig. \ref{fig:b_y-spacetime-general}. In general, simulations begun with a slanted symmetry initial condition (the `bxGauss' profile) move into the slanted symmetry cyclic state, while those began with an hourglass symmetry initial condition (the `bxSINmod' profile) settle into the odd symmetry cyclic state. Simulations started with the 
`bxASYM' profile were found to settle into the hourglass symmetry cyclic state. After $100$s of $\Omega^{-1}$, a mid-plane perturbation exits these cyclic states through an intermediate and short-lived transition state to the slanted symmetry steady state. After that, no further qualitative changes were observed. Below, we give a more detailed description of the properties of each of the four states and their behaviour.

\subsubsection{Hourglass symmetry cycles}

\begin{figure}
    \centering
    \includegraphics[width=0.5\textwidth]{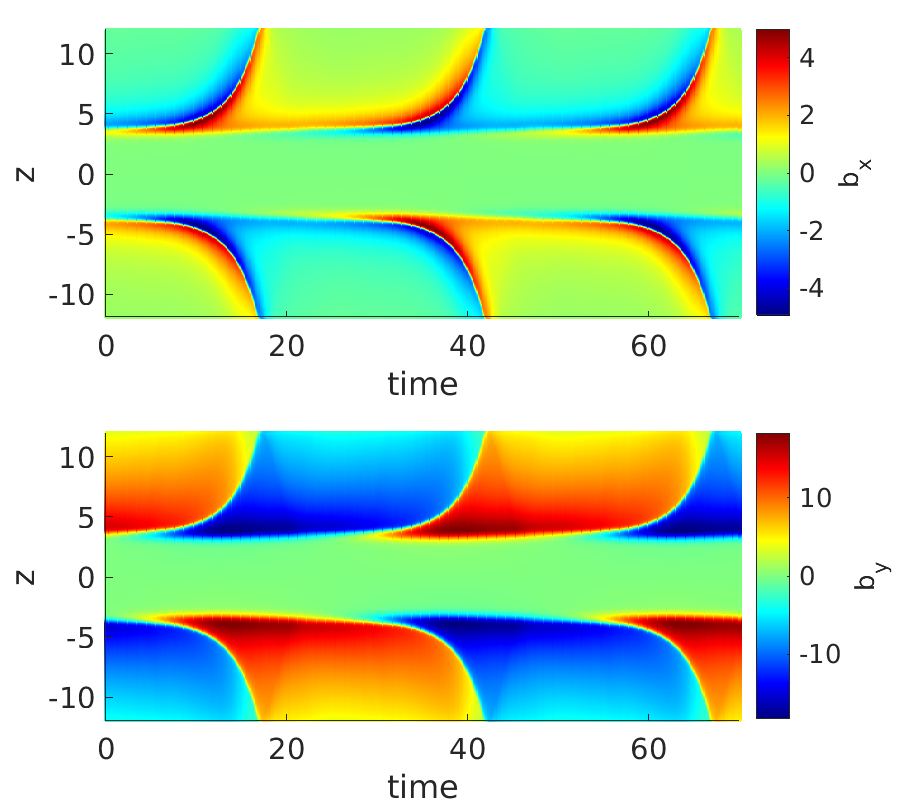}
    \caption{Space-time plot of $b_x$ (top) and $b_y$ (bottom) for the hourglass symmetry cycles observed in run \textbf{b1e5\_S}.}
    \label{fig:bt10t5_bxSINmod-bx-by}
\end{figure}

The hourglass symmetry cycles are long-lived time-dependent states where horizontal magnetic fields $b_x$ and $b_y$ have opposite signs across the mid-plane, and the horizontal velocity fields $v_x$ and $v_y$ are even in $z$. As an example, we consider here the cyclic solutions obtained for run \textbf{\text{b1e5\_S}}, with parameters $\beta_0=10^5$, $\delta=0.033$, $\Hat{z}_i=0.5$, and initiated from the `bxSINmod' profile. Fig. \ref{fig:bt10t5_bxSINmod-bx-by} shows the space-time variation of the horizontal magnetic fields $b_x$ and $b_y$, where the lower case $b$ denotes that they have been normalised with respect to the vertical field strength. Owing to the hourglass symmetric nature of the solutions, we only describe the upper half of the disc where $z > 0$.

\begin{figure}
  \centering
    \includegraphics[width=0.5\textwidth]{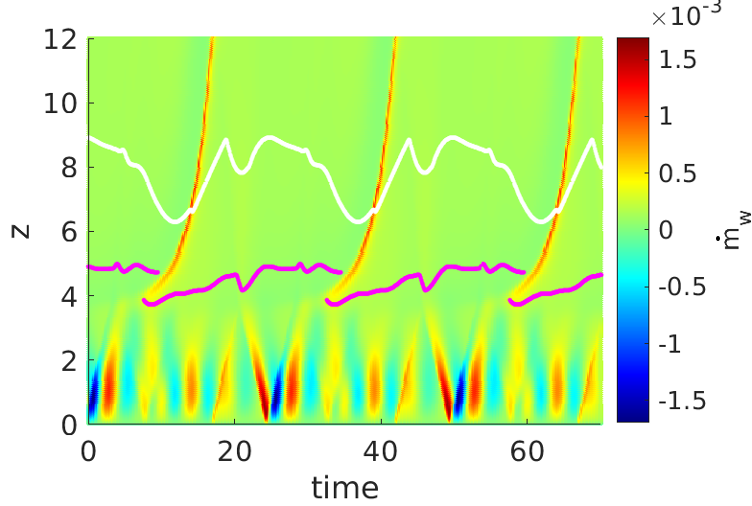}
    \caption{Space-time plot of $\Dot{m}_w$ for the hourglass symmetry cycles observed in run \textbf{b1e5\_S}. The white line indicates the sonic point, while the magenta line marks the Alfv\'en point(s).}
    \label{fig:bt10t5_bxSINmod-m_w-v_z}
\end{figure}

The period of the cycle is roughly equal to $50 \Omega^{-1}$, i.e. eight orbital periods, and is divided equally into two half-cycles where the dynamics are identical with the exception of the horizontal magnetic and velocity fields being oppositely signed. The vertical outflow, $\Dot{m}_w$, defined as the average from the boundary at one side of the disc, is not affected by this change of sign, and repeats itself every half-cycle with a period of roughly $25 \Omega^{-1}$. A strong and brief outburst roughly $6$ times the quiescent value at the boundary marks the end of each half-cycle. We observe a slow and minute sinusoidal oscillation of the mid-plane horizontal fields about $0$ with the same period as the overall cycle. The horizontal magnetic fields $b_x$ and $b_y$ are always anti-correlated with each other, and drive radial accretion or decretion flows by vertical transport of angular momentum through the $B_y B_z$ stress depending on the sign of the fields. Fig. \ref{fig:bt10t5_bxSINmod-m_w-v_z} shows the spatial variation and temporal evolution of the vertical outflow in time. The location $z_{Az}$ of the Alfv\'en point, defined as the point above which $v_z$ exceeds $v_{Az}$, fluctuates between $z = 4 H$ and $z=5 H$. It is interesting to note here that $z_{Az}$ is generally lower than the sonic point $z_s$, a consequence of the relatively weak-field regime explored in our simulations. The fast magnetosonic point is mostly approached at the simulation domain boundary, but sometimes crossing briefly occurs in the simulation domain though without significant impact, due to fluctuations in the $b_x$ and $b_y$ profiles as the outbursts pass through the atmosphere.

The quiescent stage outflow is largely steady, with $\Dot{m}_w \sim 1.4 \times 10^{-4}$ in code units. {\color{black}In the atmospheric region $z \sim 3.8 - 7.5 H$, the inclination of the poloidal field with respect to the vertical axis is significantly larger than the critical value of $30^\circ$ (which is a necessary but not sufficient criterion for a magneto-centrifugal outflow), allowing a steady wind to be launched from $z \sim 4H$ and gas to be accelerated along field lines by the magneto-centrifugal effect.} The outflow is then further enhanced by the magnetic pressure gradient in the upper atmosphere before it leaves the box.

The outburst is initiated around $z \sim 4 H$ at $t=8 \Omega^{-1}$, when $b_x$ and $b_y$ are both significantly growing in the same region and are about to reach their maximum field strengths in the half-cycle. The shape of the growing horizontal fields consists of a mid-plane region where they are flat and near zero, before developing into two peaks of opposite signs in quick succession beyond $z \sim 3 H$. The peaks of $b_x$ and $b_y$ are then accelerated upwards out of the box, with the outburst following the point where the magnetic pressure gradient is greatest between the maximum and minimum peaks of $b_x$ and $b_y$. This indicates that the gas parcel is pushed out of the box by the horizontal magnetic field peaks leaving the vertical domain. The outburst lasts for a duration of $\sim 3 \Omega^{-1}$, and at its peak has $\Dot{m}_w = 8.55 \times 10^{-4} $, up to $6$ times the quiescent value. However, the amount of gas ejected per outburst, $7 \times 10^{-4}$, is still only a tiny fraction of the overall disc mass $\Sigma=1$, and compares with $2.9\times 10^{-3}$ that is ejected over the longer quiescent interval between the outbursts. The peaks in $b_x$ and $b_y$ start decreasing in magnitude as they move beyond $z\sim 4.7 H$. At this time, a peak of the same sign slowly develops behind the lower altitude peak at around $z \sim 4 H$ which eventually becomes the higher altitude peak for the next outburst, while a peak of the opposite sign begins developing at $z \sim 3 H$, becoming the new lower altitude peak. The next half-cycle then repeats the same dynamics, except the horizontal magnetic and velocity fields have now effectively switched signs compared with the previous half-cycle.



\subsubsection{Slanted symmetry cycles}
\label{sec:unphysical_cycles}

Like the hourglass symmetry cycles, these solutions are long-lived time-dependent states, but with $b_x$ and $b_y$ being even in $z$, while $v_x$ and $v_y$ are odd. In almost all properties, these cycles are identical to the hourglass symmetry cycles, with the $b_x$ and $b_y$ having small amplitudes and a nearly flat profile in the disc region ($\lvert z \rvert < 3 H)$, while in the atmosphere, cycles of outburst up to $6$ times the mass flux of the quiescent steady outflow are driven by the same form of $b_x$ and $b_y$ peaks growing and moving up out of the vertical domain of the box. From Fig. \ref{fig:b_y-spacetime-general}, we can see that the slanted and hourglass symmetry cycles under the same simulation parameters share the same period of around $50 \Omega^{-1}$ for $\beta_0=10^5$, $\delta=0.033$, $\Hat{z}_i=0.5$, and $L_z=12$, a property which is also observed for other sets of simulation parameters. This suggests that the high diffusivity in the mid-plane region effectively cuts off magnetic communication between the upper and lower halves of the disc, and since these cycles are connected to and driven by the growth and movement of peaks in $b_x$ and $b_y$, there is (similar to \citet{BaiStone2013}) an equal chance of adopting either symmetry unless it is already set by the initial condition.

\subsubsection{Cyclic to steady wind transition state}
\label{sec:trans_description}

\begin{figure}
  \centering
    \includegraphics[width=0.5\textwidth]{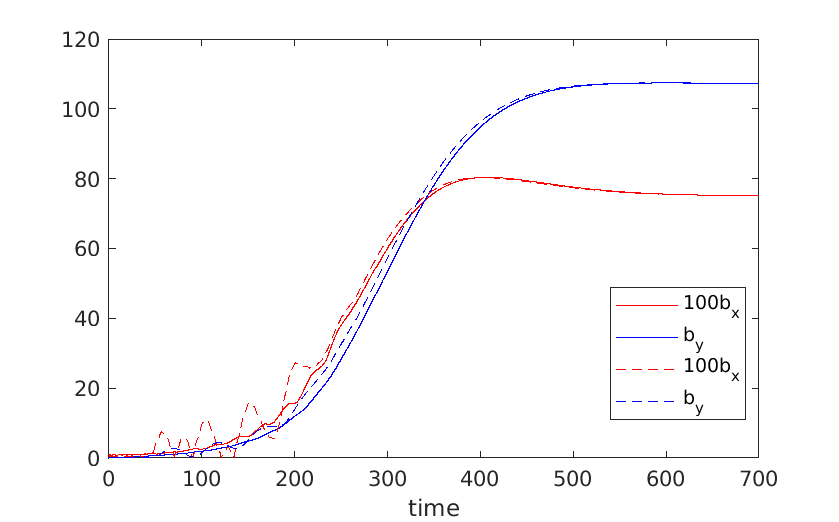}
    \caption{Plot of mid-plane $b_x$ (red) and $b_y$ (blue) against time for \textbf{b1e5\_A} (solid lines) and \textbf{b1e5\_G} (dashed lines).}
    \label{fig:Midplane_exp_plot}
\end{figure}

The hourglass and slanted symmetry cycles typically survive on a timescale of $100$s of $\Omega^{-1}$, with a weaker vertical field leading to a longer survival time. In fact, for runs initialised using 'bxSINmod' with $\beta_0=10^5$ and $10^6$, the hourglass symmetry cyclic solutions show no sign of transitioning throughout the entire runtime of the simulation up to $700 \Omega^{-1}$. Transition to the intermediate state begins with a small mid-plane bulge in $b_x$ which gets sheared into a corresponding mid-plane bulge of opposite sign in $b_y$. This bulge then grows slowly but exponentially in magnitude, as shown in Fig. \ref{fig:Midplane_exp_plot} for runs \textbf{b1e5\_A} (solid lines) and \textbf{b1e5\_G} (dashed lines). As long as the magnitude of the bulge in $b_y$ is lower than that of the $b_y$ peaks of the hourglass/slanted symmetry cycles, there is minimal effect of the growing mid-plane dynamics on the properties of the cyclic wind states, with both the magnitudes and periods of the cycles on the whole unaffected. However, once the mid-plane $b_y$ has reached the magnitude of maximum wind cycle $b_y$ peak strength, the period of the half-cycle lengthens or shortens if the sign of the mid-plane $b_y$ is of the same or opposite sign of the higher altitude $b_y$ peak respectively. The nature of the solution then changes to that of a steady wind over the next half-cycle, and the cycles stop. As in the case with the hourglass/slanted symmetry solutions, the mid-plane region disconnects the two sides of the disc, and each side of the disc effectively behaves independently from the other and interacts with mid-plane region individually.

\subsubsection{Slanted symmetric steady state}
\label{sec:unphysical_steady}

\begin{figure}
  \centering
    \includegraphics[width=0.5\textwidth]{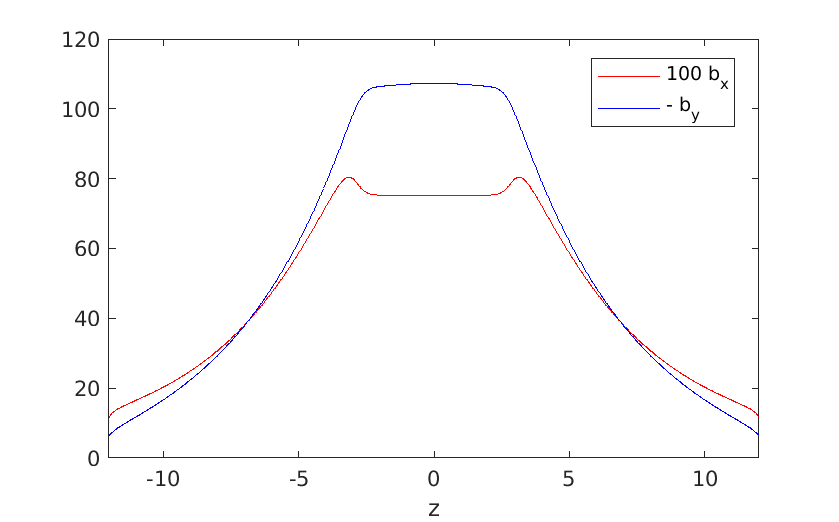}
    \caption{Vertical profiles of $b_x$ and $b_y$ for the slanted symmetry steady state with $\beta_0=10^5$.}
    \label{fig:b_x-b_y-UnphysSteady}
\end{figure}

For our simulations in Figure \ref{fig:b_y-spacetime-general} with $\beta_0=10^5$, at around $t=300 \Omega^{-1}$, the exponential growth of the mid-plane $b_x$ and $b_y$ slows down, and a steady wind solution is reached by $t=500 \Omega^{-1}$. Throughout the saturation stage, the disc has a slanted symmetry with $b_x$ and $b_y$ even in $z$ and of opposite signs, while $v_x$ and $v_y$ are odd in $z$. $b_x$ and $b_y$ have large amplitudes ($>1$) and a flat profile near the mid-plane, but $\p_z b_x, \p_z b_y \neq 0$ at the $z$ boundaries of the box. A plot showing the profile for $\beta_0=10^5$ is shown in Fig. \ref{fig:b_x-b_y-UnphysSteady}. $b_x$ has amplitude maxima in the region where the $b_x$ peaks are observed to start growing in the cyclic phase $(\lvert z \rvert \sim 3.15 H)$, while $b_y$ has its amplitude maximum at the mid-plane. The flat mid-plane profiles of $b_x$ and $b_y$ may be attributed to the large diffusivities there suppressing bending of the field lines, while the absolute strength of the magnetic field at the mid-plane corresponds to $\beta = 7.95$. $b_x$ and $b_y$ in the atmosphere always tend towards zero as $\lvert z \rvert$ increases. The Alfv\'en point occurs at around $\lvert z \rvert = 2.7 H$, while the fast magnetosonic point is approached at the simulation domain boundary but not crossed. A strong, steady and slow wind of up to 10 times the $\Dot{m}_w$ value of the quiescent state in the cyclic phase is launched. 

\subsection{Dependence on the vertical field}
\label{sec:magneisattion_effects}

We varied the vertical field strength from $\beta_0=10^2$ to $\beta_0=10^6$ and examined its effect on both the cyclic state of hourglass symmetry, and its transition to the slanted symmetry steady wind solution. 

\subsubsection{Cyclic state}

\begin{figure*}
  \begin{center}
    \includegraphics[width=0.65\columnwidth]{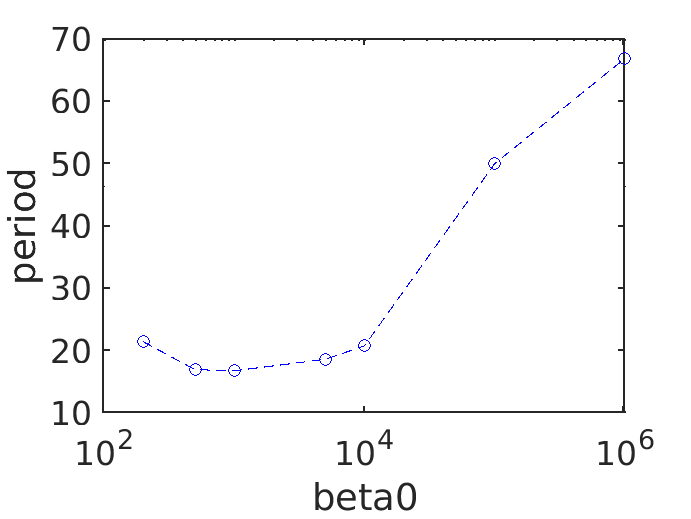}
    \includegraphics[width=0.65\columnwidth]{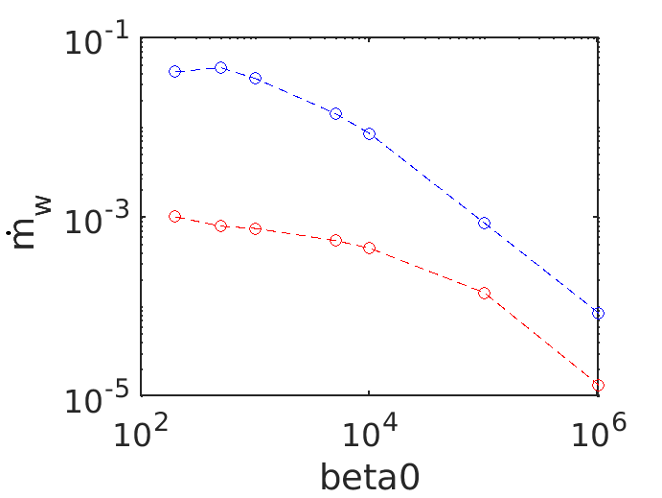}
    \includegraphics[width=0.65\columnwidth]{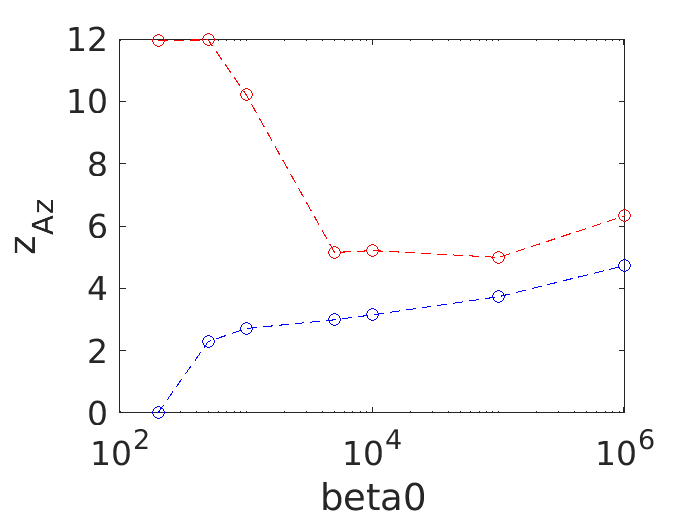}
    \caption{Variation of wind properties with $\beta_0$ for the hourglass symmetry cycles. The left panel shows the period of the cycle. The middle panel shows the average quiescent outflow strength (red) and average outburst strength (blue). The right panel shows the minimum (blue) and maximum heights (red) of the Alfv\'en point(s) in the cycle.}
    \label{fig:beta-WindProp-Cycles}
\end{center}
\end{figure*}

The variation of several key properties of the cyclic solution with $\beta_0$ are shown in Figure \ref{fig:beta-WindProp-Cycles}. For the cyclic states, as $\beta_0$ decreases, the mid-plane horizontal fields $b_{x}$ and $b_y$ become less flat, as the increased magnetisation allows a stronger current to flow there despite the higher resistivity. However, most of the bending still occurs above the region $\lvert z\rvert \sim 3H$, while the positions of the peaks as they grow are similarly located in the lower atmospheric region where $\eta$ decreases dramatically to the atmospheric value. The period of the cycles generally shortens as the magnetisation is increased, although it reaches a minimum value at around $\beta_0=500$ beyond which the period increases slightly again. Both the outburst and quiescent outflow strength increase as the magnetisation increases, and the vertical flow also becomes quicker with a lower sonic point. While the Alfv\'en point continues to vary within each half-cycle, the range of heights over which it varies stays roughly the same between $4H < \lvert z \rvert < 5H$ from $\beta_0=10^5$ to $5000$, before drastically increasing and covering the whole simulation domain by $\beta_0=200$. The height from which the outburst is launched is always located at the lower atmosphere, although it decreases from $\lvert z \rvert \sim 4.7H$ to $\lvert z \rvert \sim 2.8 H$ as the magnetisation is increased from $\beta_0 = 10^5$ to $200$.

\subsubsection{Transition state}

\begin{figure}
  \centering
    \includegraphics[width=0.5\textwidth]{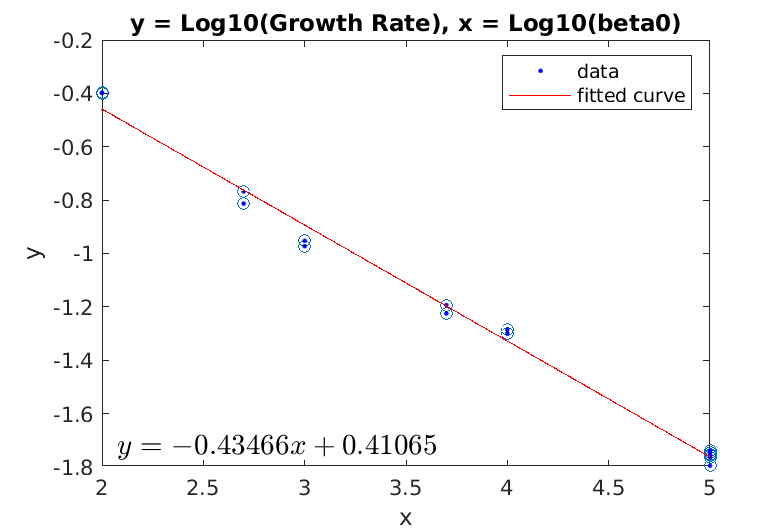}
    \caption{Plot of log10 of exponential growth rate ($y$) against log10 of $\beta_0$ ($x$) with best fit line.}
    \label{fig:sigma_v_beta0}
\end{figure}

For the transition to the slanted symmetry steady wind, an increase in field strength leads to less time spent in the cyclic state of hourglass symmetry, and a more rapid transition.  Empirically, we find that the mid-plane $b_{x}$ and $b_y$ bulge initial growth rate satisfies the relation
\begin{equation}
    \sigma \approx 10^{0.41}\cdot \beta_0^{-0.44},
\end{equation}
where $\sigma$ is the growth rate measured before saturation flattens out the exponential growth profile. Fig. \ref{fig:sigma_v_beta0} plots $\sigma$ against $\beta_0$ and the empirical fit we are able to obtain. This roughly gives us $\sigma \propto B_z$, suggesting that the mid-plane growth mechanism is magnetic in nature. The tendency for more strongly magnetised discs to more rapidly transit to the slanted symmetry steady wind solution has been previously noted in the simulations of \citet{BaiStone2013}, with the difference between our simulations being that they used more realistic diffusivity profiles, while ambipolar diffusion was also included. However, they did not examine the mechanism behind the transition, and only attributed it as possibly due to an increased difficulty in maintaining a strong current layer in the lower atmosphere (as seen in the cyclic stage with the $b_x$ and $b_y$ peaks) as the field strength is increased.

\subsubsection{Steady wind}

\begin{figure*}
  \begin{center}
    \includegraphics[width=0.65\columnwidth]{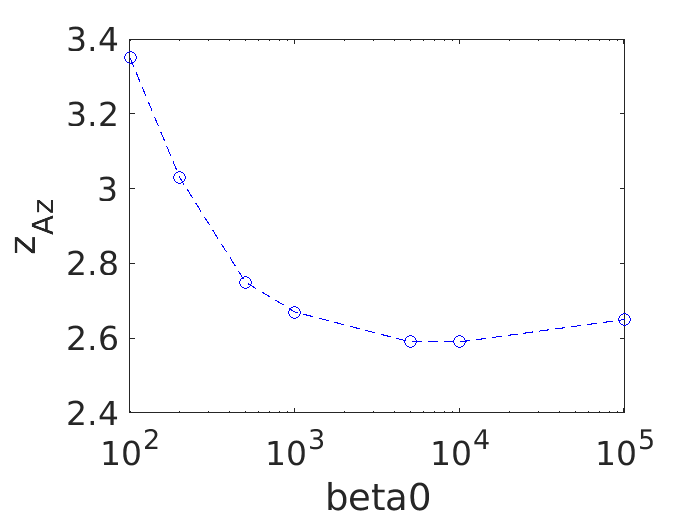}
    \includegraphics[width=0.65\columnwidth]{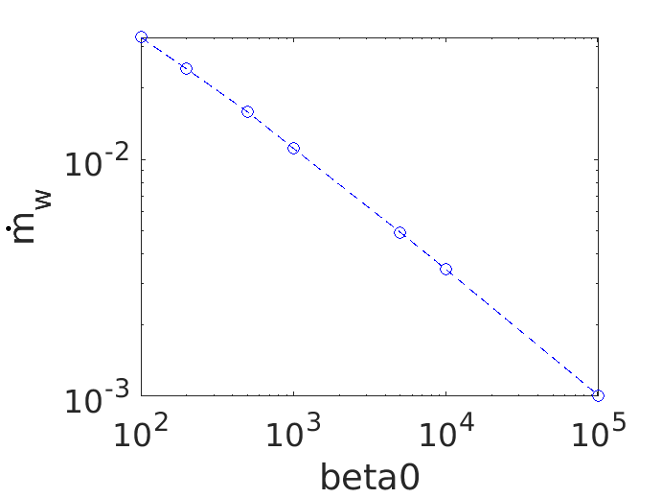}
    \includegraphics[width=0.65\columnwidth]{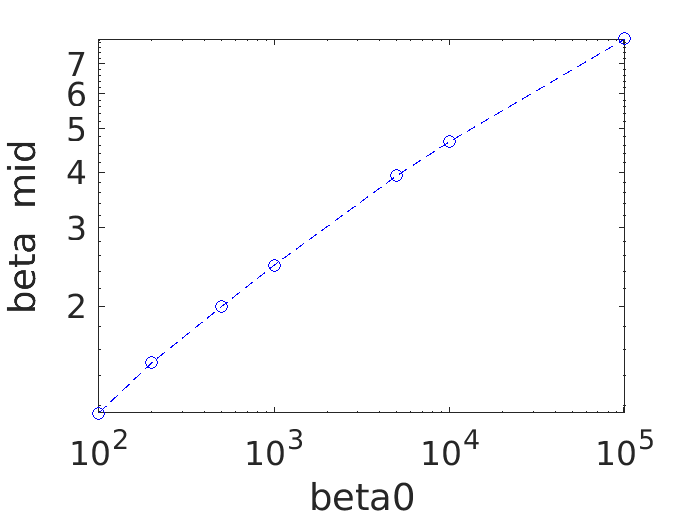}
    \caption{Variation of wind properties with $\beta_0$ for the slanted symmetry steady state. Left: position of $z_{Az}$, middle: outflow strength $\Dot{m}_w$, right: plasma $\beta$ value taking into account all three magnetic field components at the mid-plane.}
    \label{fig:beta-WindProp-UnphysSteady}
\end{center}
\end{figure*}

Finally, we examine the variation of the properties of the steady state slanted symmetry wind with disc magnetisation, which are plotted in Fig. \ref{fig:beta-WindProp-UnphysSteady}. The mass loss rate decreases with decreasing field strength and follows a power law of the form
\begin{equation}
    \Dot{m}_w \propto \beta_0^{-0.51}.
\end{equation}
This is again similar to the relation obtained in \citet{BaiStone2013} for their slanted symmetry steady winds, where the index has a value of $-0.54$. Both of these values roughly give us $\Dot{m}_w \propto B_z$, and again suggest that the vertical magnetic field still has a crucial role to play in the launching of the outflow despite being dominated by the horizontal fields in the wind-launch region. The value of the overall mid-plane $\beta$ decreases slightly from $7.95$ for $\beta_0=10^5$ to $1.15$ for $\beta_0=10^2$, but its magnitude remains of order unity. The Alfv\'en point generally falls with magnetisation, and flattens off beyond $\beta_0 = 5000$ to $z_{Az}\sim 2.6 H$.

\subsection{Robustness of the wind behaviour}

\subsubsection{Variation of box sizes}
\label{sec:box_size_effects}

\begin{table}
\caption{Comparison between $L_z=12H$ and $L_z=70H$ runs. For the cycles, the lower $\Dot{m}_w$ value is for the quiescent outflow, while the higher one is for the outburst, whereas the Alv\'en points indicate the range within which they vary.}
\centering
\label{table:Riols_SimRuns}
\begin{tabular}{l c c c c}
\hline
Property & $12H$ cycles & $70H$ cycles
& $12H$ steady & $70H$ steady\\
\hline
Period ($\Omega^{-1}$) & 50 & 57.6 & NA & NA \\
$\Dot{m}_{w}$ ($10^{-4}$) & 1.4, 8.6 & 0.48, 5.3 & 10 & 6.02\\
$z_{Az}$ ($H$) & [3.7, 5.0] & [3.6, 5.6] & 2.65 & \~ 3 \\
$b_{x,\text{max}}$ & 5.0 & 8.5 & 0.804 & 0.86\\
$b_{y,\text{max}}$ & 18 & 23 & 107 & 121\\
$\beta_\text{mid}$ & NA & NA & 7.95 & 5.54\\
\hline
\end{tabular}
\end{table}

In order to confirm that our wind behaviour is not a result of our small box size of $L_z=12H$, we also ran simulations for our $\beta_0=10^5$ simulations in boxes with $L_z=70H$. We found that the same types of wind behaviour are preserved. A comparison of the key properties between runs at the two different scale heights is listed in Table \ref{table:Riols_SimRuns}.

For the hourglass symmetry cycles, we found that both periodicity and mass loss rate converge as box size is increased. Our $L_z=70H$ cycles have a period of $57.6~\Omega^{-1}$ compared with a period of $50 ~\Omega^{-1}$ in our $L_z=12H$ runs. Both the quiescent and outburst outflow strengths are slightly weaker in the taller box, which is expected as a larger box means a greater gravitational potential for the gas to overcome to escape from the box. The $b_x$ and $b_y$ peaks in the cycles are slightly increased in magnitude as $L_z$ increases, and is probably because of the reduction of the escaping flux at the boundaries due to the smaller mass outflows \citep{Suzuki_etal_2010}. The overall cycle dynamics, including the relative positions of the wind launch point, the $b_x$ and $b_y$ peaks, and the variations of the Alfv\'en points, remain roughly the same. 

For the transition state, the mid-plane bulge exponential growth rate converges as box size is increased and is only slightly modified, with $\sigma=0.2$ for our $L_z=70H$ runs, compared with $\sigma=0.18$ for $L_z=12H$.  

The steady wind state shows similar trends in convergence to the cyclic state, with a slightly lower mass loss rate in the taller box as we would expect, and fractionally higher horizontal magnetic field strengths. Otherwise, there is no qualitative difference between the steady wind profile of the smaller box compared with the larger one.

Overall, the fact that most properties of our wind solutions were only slightly altered between our $L_z=12H$ and $70H$ simulations justifies our usage of the more computationally cost-friendly $L_z=12H$ runs to explore the parameter space, and determine the mechanisms responsible for the different types of wind behaviour we have observed. 

\subsubsection{Mass replenishment}
\label{sec:mass_repl_effects}

In order to check that the forms of the wind solution are independent of the mode of artificial mass injection, we used two different mass replenishment schemes: narrow and wide. The narrow scheme is the one used in our simulations unless otherwise stated, applying the source term of \citet{Lesur_etal_2013} and \citet{Riols_etal_2016}, as presented in equation \ref{eq:Lesur_mass_src}, with $z_i=0.1$ such that mass is injected a narrow $\lvert z \rvert < 0.1 H$ region about the mid-plane. The wide scheme, denoted `mrw', injects mass in proportion to the local density instead, and was used in the simulations of \citet{BaiStone2013}. For both injection schemes, mass replenished at each time-step is equivalent to the mass lost at the boundaries, so that the total mass of the disc is kept constant. 

We found that while the four types of wind solutions still occur when we used the `wide' scheme, there are small differences ($<10\%$) to the locations of the $b_{x,y}$ peaks in both the cyclic phases and the slanted symmetry steady state. In general, their locations are higher up in the disc, which may reflect the fact that under the `wide' scheme, the disc's density profile is more spread out than the `narrow' scheme, as mass is injected at every point rather than simply the mid-plane region. This would then imply that the locations of the growth peaks are tied to the relative strength of the vertical field to the density at that point. Another small but notable difference between the `wide' and `narrow' schemes is in the cycle dynamics. In our runs for $\beta_0=10^5$, while the $b_{x}$ and $b_y$ peaks in the `narrow' scheme are always monotonically moving away from the mid-plane, the peaks in the `wide' scheme have a brief period of small oscillations of its position in the region $3H <\lvert z\rvert < 4.75 H$ during which its growth rate also decreases and increases, before the same rapid acceleration out of the box occurs once they pass beyond $\lvert z \rvert \sim 4.75H $. Again, we attribute this difference to the fact that the `wide' scheme artificially changes the density profile across the whole disc, and points to the sensitivity of the cycle mechanism to the density profile in the $3H <\lvert z\rvert < 4.75 H$ region as the reason for the small oscillation in the $b_{x}$ and $b_y$ peaks' position. Both the mass loss rate and periodicity also only slightly altered by the `wide' scheme and its effect is not significant.

To see whether the outburst behaviour is linked to the sudden increase in mass replenishment at those times, we did a run for our $\beta_0=10^5$ simulation in the cycle phase, where we set the mass replenishment to be constant in time instead. We found that cycle dynamics is unaffected by this change, which is not surprising given that even though the outbursts have significantly higher mass loss rates than the quiescent stage, they are still small when integrated in time compared with the total disc mass. We also did a few runs where there is no mass replenishment at all, and found the cycles and periodic outburst behaviour to still persist in the absence of mass injection, and as long as the overall mass loss is not significant, there is no notable quantitative difference between the solutions. 

\subsubsection{Ideal MHD in the atmosphere}

One caveat in our model with regard to mimicking real protoplanetary discs is in the diffusivity profile used. In particular, for most of our runs we lower the Ohmic diffusivity to be $0.5\%$ that of the mid-plane value in the atmosphere, whereas one might argue that it would be more realistic to have ideal MHD due to the high FUV ionisation there. To test whether an ideal MHD atmosphere would make a difference to our results, we conducted four runs for our $\beta_0=10^5$ simulations, initialised from each of the four wind solution states, but with $\eta_\infty$ set to $0$. 

We found that while the general dynamics of the cyclic states is not changed, the peaks in $b_x$ and $b_y$ become more pronounced, with $b_{x,\text{max}} = 11.3$ and $b_{y,\text{max}} = 23.2$ compared with $b_{x,\text{max}} = 5.0$ and $b_{y,\text{max}} = 18.0$ when $\eta_\infty=0.005\eta_0$. The period of the cycles also becomes shorter, with $T = 20 \Omega^{-1}$ instead of $T=50\Omega^{-1}$ previously. The range of heights through which the Alfv\'en point moves also becomes lower, from $4H<\lvert z\rvert <5.5H$ to $2.89H<\lvert z\rvert <4.11H$, and a lower height above which the peaks will be significantly accelerated up out of the disc. The outburst becomes about $3$ times stronger than in the more diffusive case, corresponding to the greater density of the lower launch point in the disc.

For our slanted symmetry steady state run, we observe almost no quantitative difference for the background steady state when $\eta_\infty=0$. A very small (period of $73 \Omega^{-1}$) perturbation in $b_x$ sometimes occurs near the twin peaks at $\lvert z \rvert \sim 3H$, which gets rapdily advected upwards out of the disc, but is generally negligible compared with the profile. 

Overall, this points to our simulation runs with the more diffusive atmosphere as still being able to capture the essential behaviour of the wind solutions as we would expect from the more realistic ideal MHD atmosphere. The more enhanced peaks in the ideal MHD atmosphere runs point towards the sensitivity of both particularly the mechanism behind the cyclic state to the resistivity profile of the disc, an effect which will be explored in greater detail in section \ref{sec:cycles-interpretation}.

\subsubsection{Half disc simulations}

We conducted a number of half disc simulations with $0<z<L_z$ only where we enforced the traditional hourglass symmetry through equatorial symmetry conditions at the mid-plane, with $\rho(-z)=\rho(z)$, $v_{x,y}(-z)=v_{x,y}(z)$, $v_z(-z)=-v_z(z)$, $B_{x,y}(-z)=-B_{x,y}(z)$ and $B_z(-z)=B_z(z)$. Unsurprisingly, only hourglass symmetry cycles were recovered in this regime, with the exact same properties as the ones in our full disc simulations. In cases where the growing mid-plane bulge rapidly disrupts the cyclic stage, we used data from these half disc simulations to analyse the behaviour of the cyclic phase.

\section{Investigation of the wind cycle mechanism}
\label{sec:cycles-interpretation}

In light of the various types of wind solutions recovered in our simulations, there are several questions we would like to address: What is the mechanism behind the wind cycles? What causes the transition from a cyclic wind to a steady one? Why is there a mid-plane bulge in $b_x$ and $b_y$ that grows exponentially, and what causes it to saturate in the slanted symmetry steady state? We begin in this section by investigating the wind cycle mechanism, while Section \ref{sec:trans} discusses the transition from cycles to steady wind, and Section \ref{sec:bulge-sat} addresses the growing mid-plane bulge and its saturation.

We present here a more detailed description and interpretation of the cyclic solutions based on the hourglass symmetry run obtained for $\beta_0=10^5$, $\delta = 0.033$, $z_i = 0.5 H$. However, it should be noted that the same dynamics is also present across the cyclic solutions, and that the same mechanism is at work.

\begin{figure}
  \centering
    \includegraphics[width=0.5\textwidth]{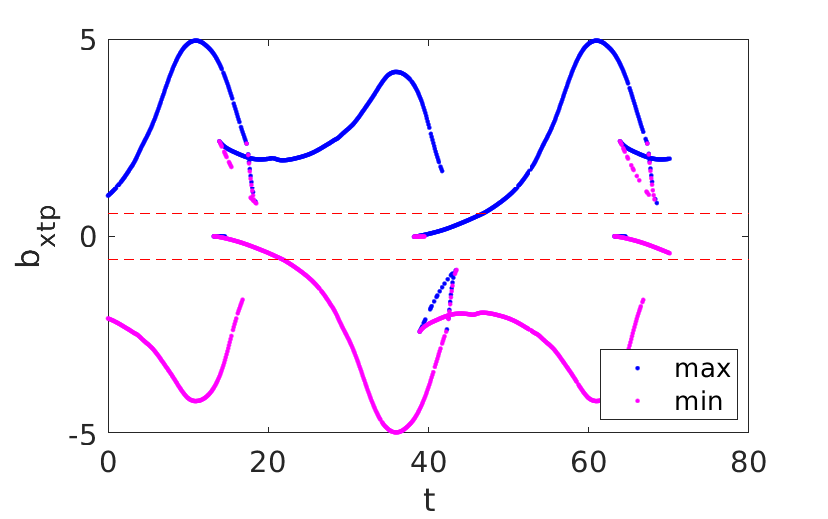}
    \caption{Plot of $b_x$ peaks strength against time. The blue line is for peaks that are maxima, while the magenta line is for peaks that are minima, and the analysis is done only for the upper half of the disc.}
    \label{fig:bxtp-t}
\end{figure}

First, we analyse the region $ 3H < \lvert z \rvert < 4H$, where new $b_x$ and $b_y$ peaks are observed to grow at the beginning of each cycle. We hypothesise that this growth is a manifestation of an MRI mode, which becomes active in this region. It is a well known result that the MRI is largely suppressed by Ohmic diffusion when the Elsasser number
\begin{equation}
    \Lambda \equiv \f{v_{Az}^2}{\eta \Omega }
\end{equation}
is smaller than $1$ \citep{SanoMiyama_1999}. 
In the mid-plane region under the resistivity profile we have chosen, $\Lambda$ at the mid-plane is of order $10^{-5}$, and increases to only $10^{-3}$ at $z=2H$, far too small for the MRI to be active. However, this changes dramatically at around $z\sim 3H$, where the diffusivity is rapidly reduced to its atmospheric value, coupled with a rapid decrease of the local density. At $z=3H$, we have $\Lambda\approx0.04$, but by $z=3.9H$, $\Lambda$ has reached $1$, and continues to increase with height. We should therefore expect the MRI to cause growth of $b_x$ and $b_y$ as $\lvert z\rvert $ approaches $3.9H$, and a significant increase in growth rate when $\lvert z\rvert$ surpasses it, which is indeed what we observe in the behaviour of the peaks. { \color{black}We identify the relevant MRI modes as those with vertical mode number $n=2$ or $3$, in which the profiles of $b_x$ and $b_y$ each have a single node on each side of the mid-plane, the $n=3$ mode also having a node at the mid-plane. These modes are usually discussed in the ideal MHD context, but given the mid-plane region is highly resistive, the node in the mid-plane for $n=3$ is of less importance, because the mode is largely suppressed in this region. }The high mid-plane resistivity effectively shuts down communication between the two sides of the disc for this mode, allowing each side to have the further from mid-plane peak as either positive or negative, depending on the history of the half-disc profile. This may explain why the hourglass and slanted symmetry cycles share the same periodicity, as the MRI-dead mid-plane causes neighbouring modes of opposite symmetry (in particular the $n=2$ and $n=3$ modes) to become degenerate and share the same growth rate, and also to have the same eigenfunction and share the same mode shape, with the exception of the overall symmetry about the mid-plane. Fig. \ref{fig:bxtp-t} shows how the peaks of $b_x$, both primary and secondary, grow with time in the upper half of the disc over around one and a half cycles. By applying fits, we verified that the initial growth of these peaks is indeed exponential, with a measured growth rate $\sigma = 0.18$. However, as the peaks themselves rapidly reach saturation in the non-linear regime and are of the same order of magnitudue as the background, we do not expect the modes to be recovered in a linear mode analysis, which ignores time-dependent terms and assumes a steady background.



\begin{figure}
  \centering
    \includegraphics[width=0.5\textwidth]{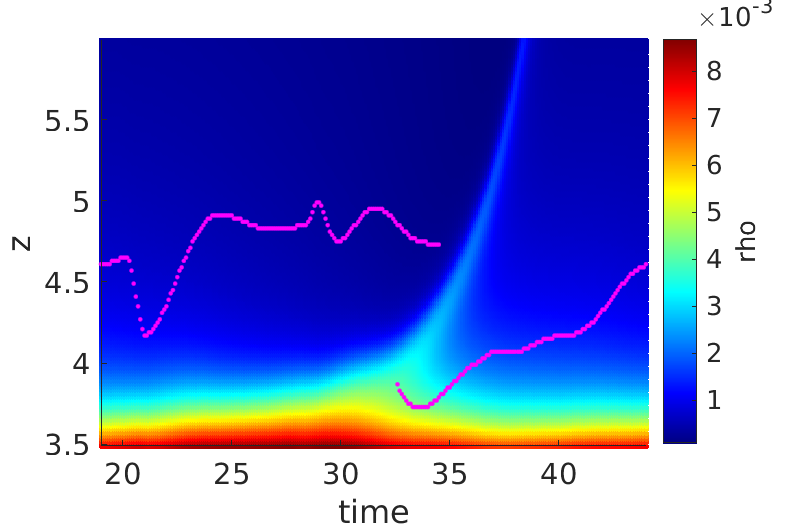}
    \includegraphics[width=0.5\textwidth]{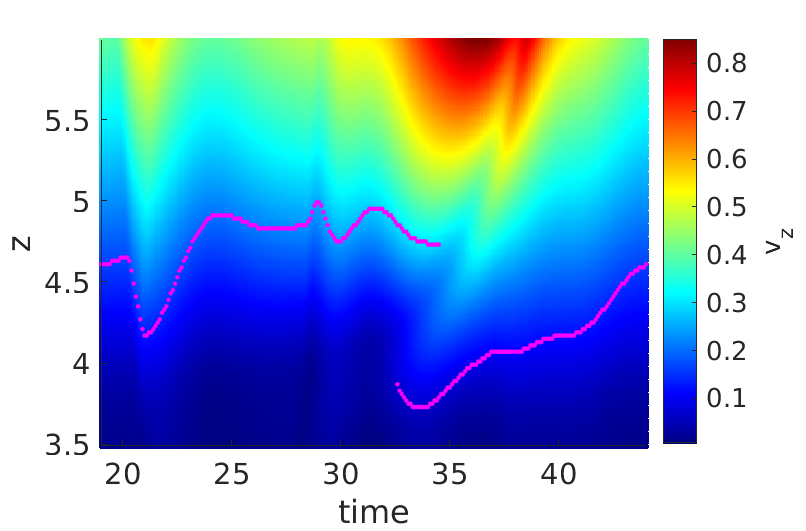}
    \includegraphics[width=0.5\textwidth]{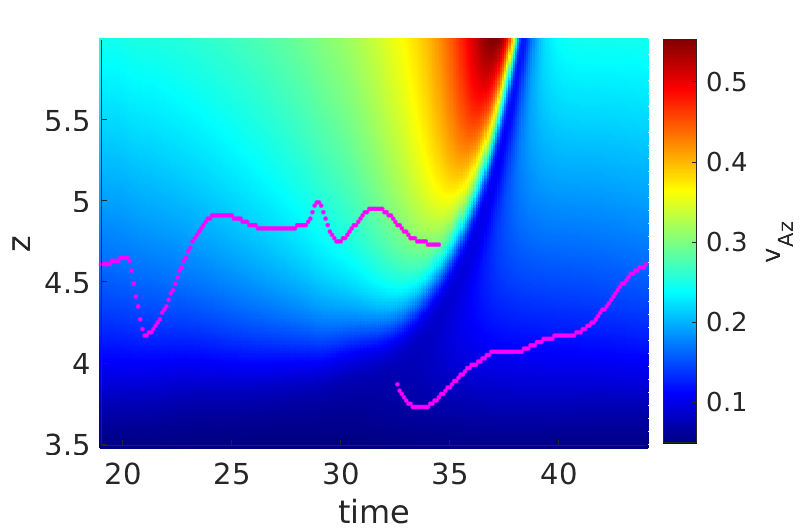}
    \caption{Space-time plots zooming into the MRI-wind region in run \textbf{b1e5\_S} over one half-cycle. The magenta lines plot the Alfv\'en point(s).}
    \label{fig:PhysCycles-Spacetime-MRI-wind-region}
\end{figure}

Having established that the growth of the $b_x$ and $b_y$ peaks is mostly due to the $n=2$ or $3$ MRI mode, we now turn to examine what contributes to their saturation and eventual acceleration up out of the disc. The Alfv\'en point marks the height above which vertical advection dominates over MRI dynamics. We would therefore expect MRI modes excited above the Alfv\'en point to be rapidly advected out of the disc, preventing further growth. Figure \ref{fig:PhysCycles-Spacetime-MRI-wind-region} gives detailed space-time plots of the density (top), vertical velocity (middle) and vertical Alfv\'en velocity (bottom) over one half-cycle, with the Alfv\'en point(s) marked with magenta dots. At the beginning of a half-cycle (which we define as after the previous outburst has been clearly emitted from the disc surface), the Alfv\'en point is at around $\lvert z\rvert = 4.5H$. A slow wind is present upwards of $\lvert z\rvert \sim 4.5H$, which is driven by the gradually weakening but nevertheless significant magnetic pressure gradient from $b_y$ of the previous half-cycle. This weakening magnetic pressure gradient correspondingly leads to a lower vertical velocity in the slow wind region, and an overall small increase in height of the vertical Alfv\'en point. 

\begin{figure}
  \centering
    \includegraphics[width=0.5\textwidth]{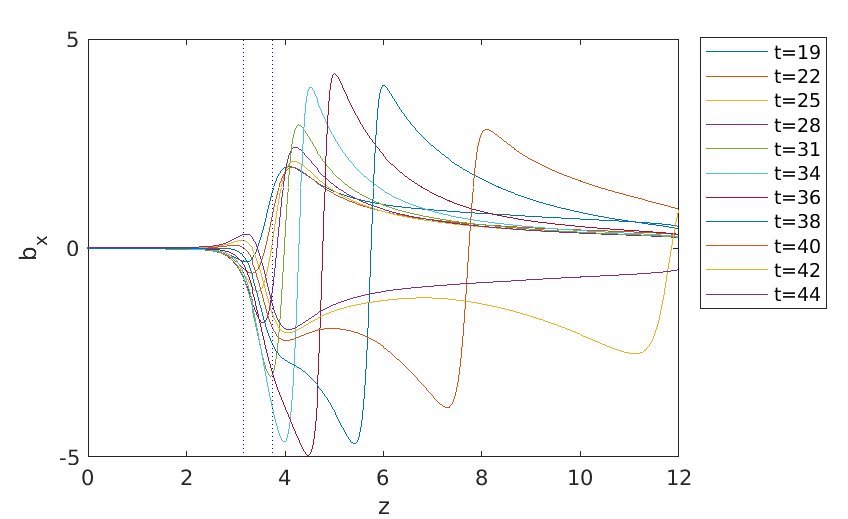}
    \includegraphics[width=0.5\textwidth]{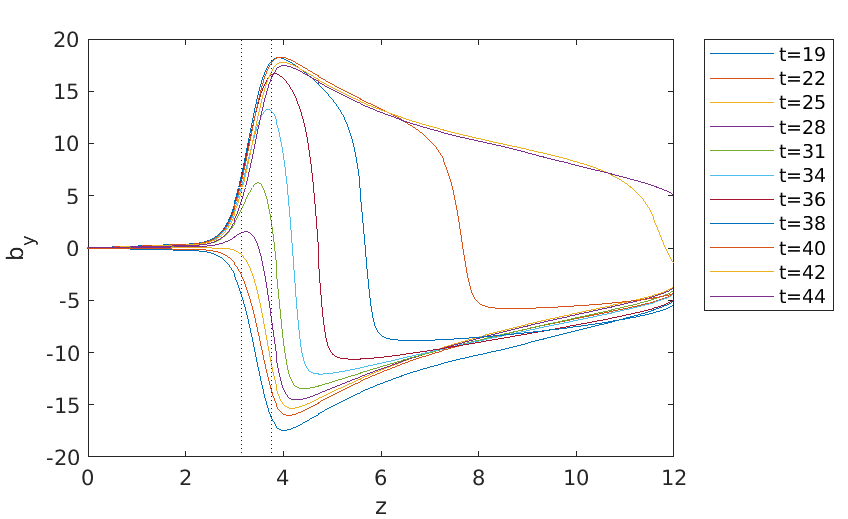}
    \includegraphics[width=0.5\textwidth]{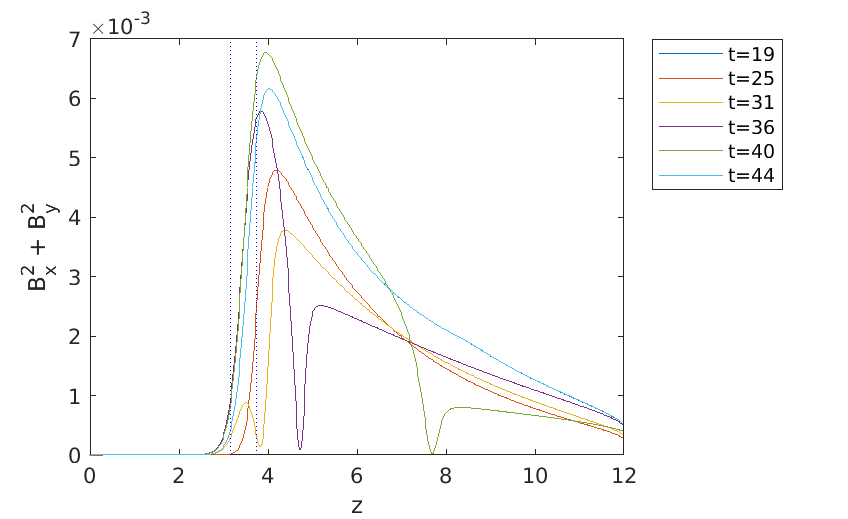}
    \caption{Time snapshots of $b_x$ (top), $b_y$ (middle) and the horizontal magnetic pressure $B_x^2+B_y^2$ (bottom)
    in run \textbf{b1e5\_S} over one half-cycle. The two blue dotted lines in each plot mark the locations $z=3.15H$ and $z=3.75H$.}
    \label{fig:PhysCycles-b_x-b_y-tsnaps}
\end{figure}

As this is happening, the $n=2$ or $3$ mode is active at a lower height of around $z\sim 3.15 H$, with a primary peak in $b_x$ beginning to grow there of opposite polarity to the $b_x$ profile in the upper atmosphere, which becomes the secondary peak of the mode. The corresponding $b_y$ of opposite sign is generated through shearing of $b_x$. Fig. \ref{fig:PhysCycles-b_x-b_y-tsnaps} plots snapshots of the $b_x$ and $b_y$ profiles over the half-cycle, as well as the horizontal magnetic pressure proportional to $B_x^2+B_y^2$. At the same time, the disc undergoes a slow expansion of its density profile, which we attribute to the disc moving back to hydrostatic equilibrium, having lost significant mass from the $\lvert z\rvert \sim 4H$ region in the previous outburst.  
This expansion slowly pushes the MRI mode further upwards into the atmosphere, while the background vertical velocity remains roughly constant. As the $b_x$ and $b_y$ peaks grow however, the node between adjacent peaks results in a magnetic pressure trap that begins to confine gas from the upper layers of the disc and move them higher up with the mode into the atmosphere. Eventually, by about $2/3$ of the way into the half-cycle, the increase in density in the lower atmosphere decreases the vertical Alfv\'en speed there so much that a second Alfv\'en point forms at a lower altitude of $z\sim 3.75 H$ below the $b_x$ and $b_y$ peaks. As a result, advection now dominates the mode dynamics, accelerating the peaks upwards into the upper atmosphere and stopping their growth. As the MRI peaks are accelerated upwards, the large magnetic pressure dip between them continues to trap gas in that region, and moves it upwards out of the disc with the peaks. This then forms the outburst gas parcel that marks the end of the half-cycle as it leaves the simulation domain. Finally, with the loss of the gas parcel, the overall density profile is reduced back to the more compact state at the start. The half-cycle then repeats itself with the horizontal field variables taking values of the opposite polarity, and the old primary peak profile becomes the secondary peak of the new half-cycle.

\subsection{Changes in the dynamics with increasing magnetisation}

Here, we explain the changes in cycle dynamics with disc magnetisation described in Section \ref{sec:magneisattion_effects} using our mechanism. Even for the strongest field strength we used of $\beta_0=200$, the Ohmic Elsasser number $\Lambda$ is still very much $<1$ in the mid-plane region, and only reaches the critical value of $1$ for $b_x$ and $b_y$ peak growth at $\lvert z \rvert = 3.27H$. Hence the two sides of the disc are still `disconnected' from each other concerning the $n=2$ or $3$ MRI modes, and the peak growth mechanism driving the cycle dynamics happen continues to occur in lower atmosphere. The period of the cycles is tied to how rapidly the $n=2$ or $3$ mode peaks grow sufficiently to trap gas in the disc surface layers and move them upwards to cause the occurrence of the second Alfv\'en point. Given a stronger vertical field, we would expect the growth rate of the mode to increase as long as the field is not so strong that the MRI is suppressed \citep{Latter_etal_2010}. Hence it is not surprising that the period of the cycles decreases as the magnetisation increases. The stronger vertical field also lowers the region in which $\Lambda > 1$, allowing the $n=2$ or $3$ mode peaks to develop lower in the disc where the density is higher, resulting in a stronger outflow in the outbursts. The outflow in the quiescent stage is enhanced slightly by the stronger magnetisation, as the horizontal fields are stronger and therefore can produce a steeper magnetic pressure gradient. Perhaps the greatest change to the cycle dynamics, as seen in Fig. \ref{fig:beta-WindProp-Cycles}, is the range of heights that the Alfv\'en point traverses as the magnetisation is increased. We will address the issue of the maximum height the Alfv\'en point reaches in the next subsection, but we confirmed that the same mechanism is indeed at work in driving the cycles by observing that the outbursts are launched at the times when the second Alfv\'en point appears. The position of this Alfv\'en point does not necessarily matter, as long as it is lower than the $b_x$ and $b_y$ peaks, which is the case for all our simulations. 

\subsection{The absence of higher order modes}

One question concerning our explanation of the cyclic state mechanism is why we only see the excitation of the $n=2$ or $3$ mode, while higher order modes are absent. For discs with $\beta_0 > 10^3$, we hypothesise that it is due to the generally low height ($\lvert z \rvert < 6H$) that the Alfv\'en point reaches even at its maximum in the cycle, thus higher order modes with multiple peaks, some of which would be located above this height, are advected rapidly out of the disc before any significant development. We confirmed this theory by repeating our runs in this regime from the cyclic state but with $v_z$ arbitrarily set to $0$ at each time-step. We see the rapid development of modes with multiple peaks in $b_x$ and $b_y$ in the upper atmosphere not present before, which quickly outgrow the original $n=2$ or $3$ mode peaks in the lower atmosphere. For discs with lower $\beta_0$, on the other hand, the Alfv\'en point varies over a much wider range, and reaches the box boundary and beyond for significant parts of each half-cycle. In these cases, we attribute the lack of higher order modes to the fact that they are shut down by the higher magnetisation, as seen in figure 2 of \citet{Lesur_etal_2013}. We tested this hypothesis by again setting $v_z$ to $0$ for discs in the low $\beta_0$ regime, and confirming that the $n=2$ or $3$ modes in these cases are indeed the fastest growing modes. 

\subsection{Comparison with the cycle dynamics of \citet{Riols_etal_2016}}
\label{sec:RiolsComp}

In the vertical 1D MHD simulations of \citet{Riols_etal_2016}, they also observed the formation of wind cycles. These were mostly done in the ideal MHD regime, but were shown to be robust even in the presence of Ohmic resistivity. Here we would like to examine the differences between their work and ours, and why our wind cycle mechanism is distinct from the one proposed by \citet{Riols_etal_2016}.

First, we note the very different magnetisation regimes that are considered in our papers. While their work focuses on a narrow range of strongly magnetised discs with $2.51 <\beta < 16$, ours explores a much more weakly magnetised regime of $10^2<\beta < 10^6$. The corresponding strengths in the horizontal magnetic fields $B_x$ and $B_y$ mean that their discs are much more significantly compressed in certain phases of the cycle than ours, as indicated by the middle and bottom panels of their Fig. 2. Consequently, while compression of the disc by the growing magnetic perturbations is the major cause of the shutting down of the MRI modes in their paper, our discs are still expanding when the MRI mode stops growing and is advected out of the disc. In a way, the MRI in our discs never truly shuts down, but rather, as one mode is advected out of the disc due to having crossed the Alfv\'en point, a new one develops in its place at a lower altitude and with the opposite polarity. Often this happens at the same time as the mode advection, hence making it difficult in our case to define when exactly a half-cycle ends or begins. In contrast, the cycles of $b_x$ and $b_y$ in the \citet{Riols_etal_2016} paper are always well separated in time, and the modes preserve the same sign across cycles. There is also a significant phase shift in time between the $b_x$ and $b_y$ maxima in the Riols cycles, whereas ours are always in phase. 

Second, the nature of the outbursts themselves is significantly different. While ours are due to material trapped by the peaks of the magnetic perturbation being advected of the disc, forming a short, concentrated burst, theirs involves expansion of the disc atmosphere over a longer timescale pushing material out of the disc, forming a more spread out wind maximum. 

Third, even though \citet{Riols_etal_2016} ran simulations with a resistive background, the values they used correspond to a minimum $\Lambda$ of $1.7$, which is not sufficient to significantly suppress the MRI. They also used a uniform diffusivity profile, and so would not have the situation as we do of a mid-plane region that effectively cuts off communication between the two sides of the disc, at least concerning the cycle dynamics.

\begin{figure}
  \centering
    \includegraphics[width=0.5\textwidth]{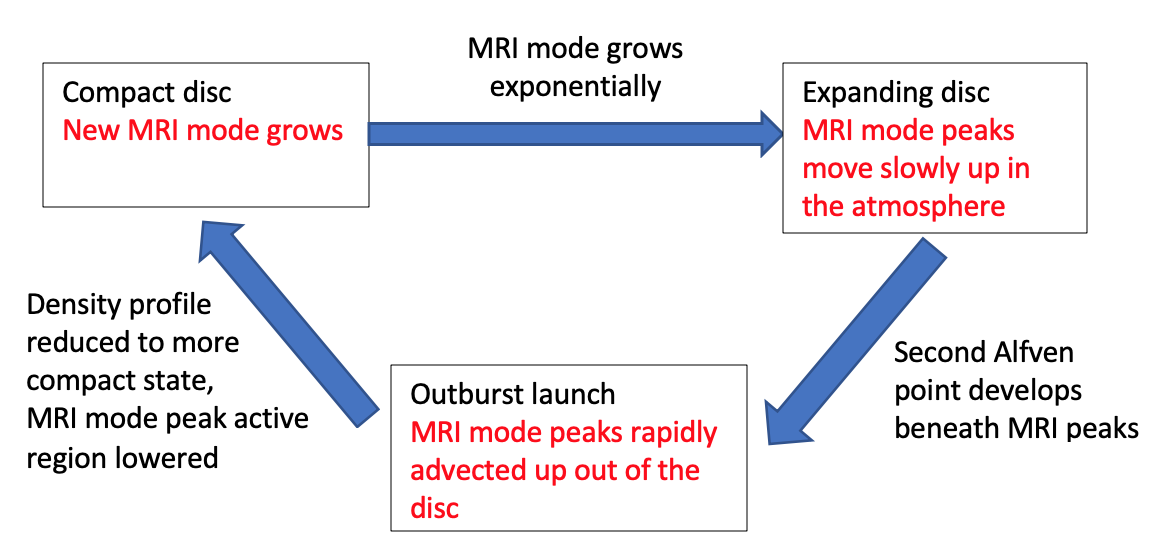}
    \caption{A sketch of the proposed cycle mechanism. To be compared with figure 14 of \citet{Riols_etal_2016}.}
    \label{fig:CyclesScheme}
\end{figure}

To summarise, the cycles we recover here are significantly different from the ones found by \citet{Riols_etal_2016}. Rather than compression-driven as in the Riols cycles, where a strong magnetic compression shuts down the MRI and its weakening then allows the disc to be MRI active again, our periodic cycles are rather advection-driven, where the rapid advecting of MRI modes out of the disc is the mechanism that prevents its further growth, and the weakening of this advection allows the growth of the mode for the next half-cycle. Figure \ref{fig:CyclesScheme}, which gives a sketch of our cycle mechanism, should be compared with figure 14 of \citet{Riols_etal_2016} to illustrate the differences between our cycles.

\section{Mechanism for the transition to a steady wind of slanted symmetry}
\label{sec:trans}

The aim of this section is to explain what causes the transition from a cyclic wind to a steady one of slanted symmetry. We do not address the origin of the mid-plane bulge itself, which we will examine in detail in the next section. The key questions we would like to answer are: Why does the final half-cycle where the horizontal fields of the atmosphere are of the same sign as the mid-plane bulge lengthen? What is the dynamics of the final shortened half-cycle and of the shutdown of the cycles?

We first recall our conclusion from the previous section that the cycles are primarily driven by the MRI combined with vertical advection, and that they depend on a delicate arrangement of the relative positions of the Alfv\'en points and the peaks of the fast growing $n=2$ or $3$ MRI mode in the atmospheric region. As noted in Section \ref{sec:trans_description}, the cycles pretty much continue as before until the magnitude of $b_y$ of the growing mid-plane bulge is comparable to the maximum magnitude that is observed in the $b_y$ peaks of the cycles. We now expand on why this is indeed a significant turning point in the disc dynamics from cycles to a steady wind, and how this transition occurs. To illustrate the dynamics of this process, we focus on the transition as observed in the run \textbf{b1e5\_G}, where $\beta_0=10^5$, $L_z=12H$ and $\delta = 0.033$. The transition is from the slanted symmetry cyclic state to the slanted symmetry steady wind, but the same mechanism can also be individually applied to each half of the discs transiting from the hourglass symmetry cycles.

\subsection{Lengthening of the final half-cycle of the same sign}

\begin{figure}
  \centering
    \includegraphics[width=0.5\textwidth]{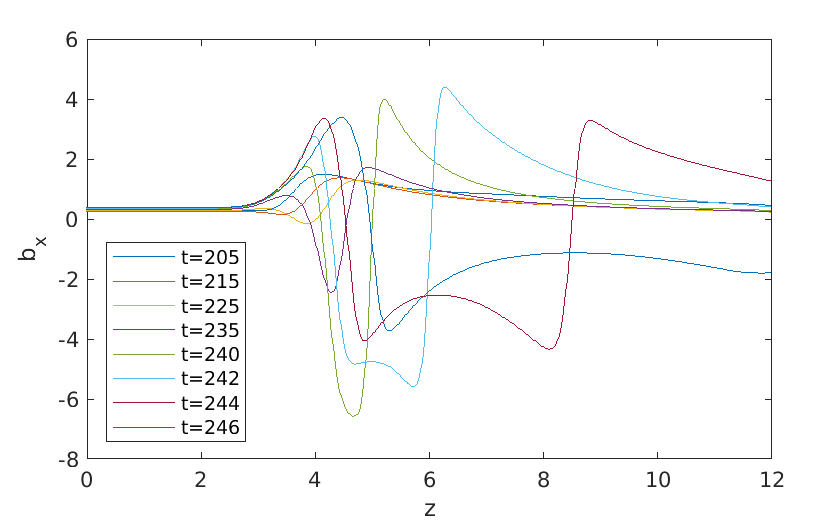}
    \includegraphics[width=0.5\textwidth]{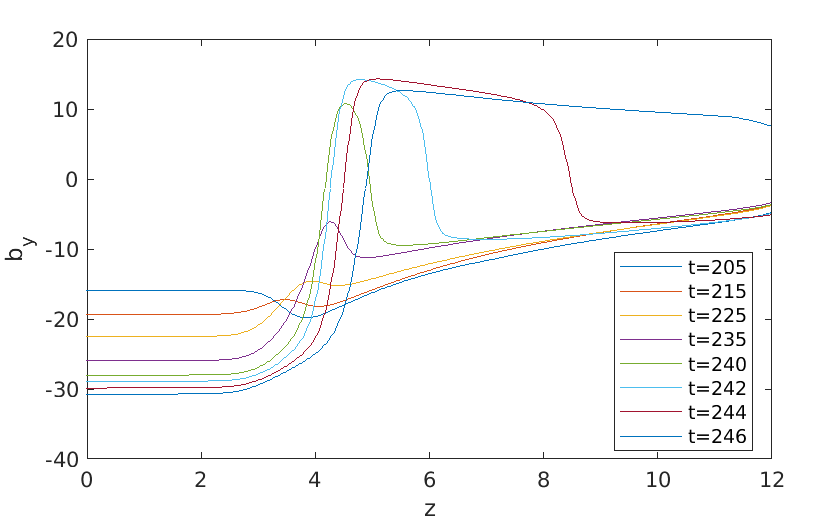}
    \includegraphics[width=0.5\textwidth]{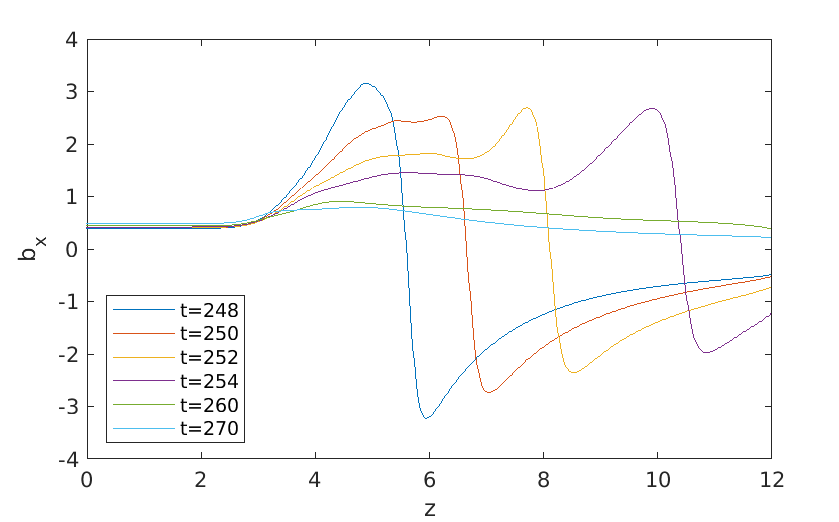}
    \includegraphics[width=0.5\textwidth]{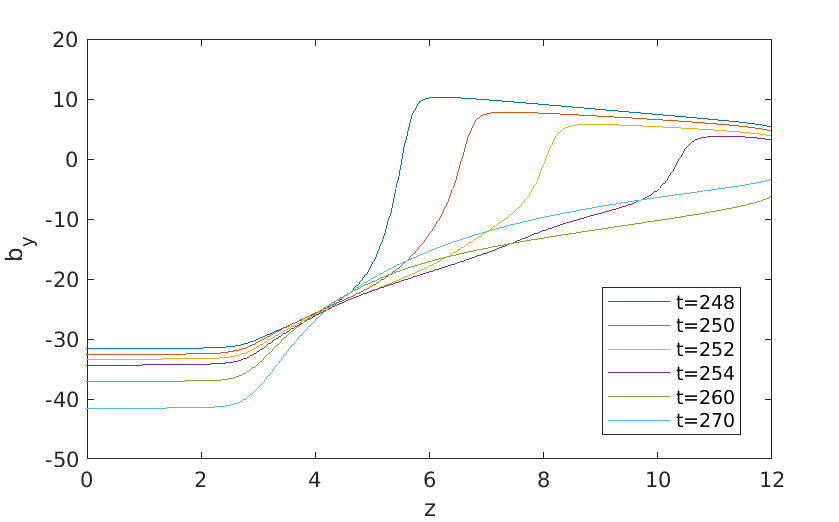}
    \caption{$b_x$ and $b_y$ time snapshots over the transition in run \textbf{b1e5\_G}. The top two plots depict the lengthening of the final half-cycle of the same sign as the bulge, while the bottom two plots depict the shutdown of the cycles.}
    \label{fig:Trans-b_x-b_y-tsnaps}
\end{figure}

First, we address the penultimate half-cycle where $b_x$ and $b_y$ begin with the same sign as the mid-plane bulge at $t=205 ~\Omega^{-1}$. The top two panels of Fig. \ref{fig:Trans-b_x-b_y-tsnaps} shows time snapshots of the $b_x$ and $b_y$ profiles respectively. As in a normal half-cycle, new $b_x$ and $b_y$ peaks of opposite sign to the current state grow in the $\lvert z \rvert \sim 3.5H$ region of the more compact disc, becoming more visible from $t=215$ to $t=235$. At this stage in the cycle, we expect the disc to be expanding slowly from its more compact form by the gas pressure gradient, having lost mass in the wind launch region from the previous outburst. However, this time, as the peaks need to be connected to the mid-plane bulge, an additional magnetic pressure gradient, particularly from $b_y$, is formed at $\lvert z \rvert \sim 3.5H$, which expands a much larger portion of the disc with the growing mode. The disc, on the other hand, is prevented from just spreading out into the atmosphere by another magnetic pressure gradient with opposite sign just below the new $b_x$ and $b_y$ primary peaks. This results in a much more expanded disc than before, as can be seen by comparing the first panel of figure \ref{fig:PhysCycles-Spacetime-MRI-wind-region} and figure \ref{fig:Transition-Spacetime-MRI-wind-region}. The dramatic increase in the height of the disc surface (which we define to be where $\rho = 5 \times 10^{-3}$) leads to a much higher Alfv\'en point in the atmosphere despite its slow increase with respect with the disc surface as in a normal half-cycle. This is further enhanced by the extended time it takes for the new $n=2$ or $3$ peaks of opposite sign at the disc surface to gain sufficient strength to trap and move the gas parcel of the upper disc layers to cause the formation of the second Alfv\'en point beneath the peaks, as the mode peaks have to overcome the initial bias of opposing sign due to the mid-plane bulge. As a result, the half-cycle is lengthened, before the same outburst behaviour as the cyclic stage occurs due to advection once the second Alfv\'en point forms beneath the mode peaks.

\subsection{Shutdown of the cycles}

\begin{figure}
  \centering
    \includegraphics[width=0.5\textwidth]{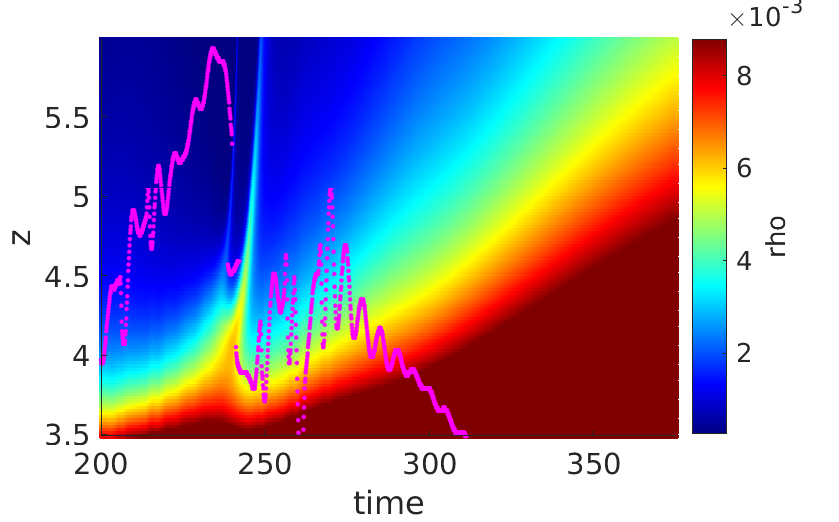}
    \caption{Space-time plot of density zooming into the MRI-wind region in run \textbf{b1e5\_G}. The magenta line depicts the Alfv\'en point.}
    \label{fig:Transition-Spacetime-MRI-wind-region}
\end{figure}

Having addressed the lengthening of the second last half-cycle, we now turn to the final shortened half-cycle and the shutdown of the cycles. As the disc returns to the more compact state at $t=248 \Omega^{-1}$, the new $b_y$ peak that develops is now of the same sign as the mid-plane bulge, but is completely dwarfed by the mid-plane bulge strength. The $b_x$ and $b_y$ peaks fail to develop sufficiently to cause a node to appear between the primary and secondary peaks. As a result, no magnetic pressure barrier develops to keep the disc from spreading out without check into the atmosphere. At the same time, a large magnetic pressure gradient, particularly from $B_y$ due to connection of the profile with the now overwhelming mid-plane bulge, pushes gas in the disc upwards, dramatically altering the density profile as can be seen from $t=248\Omega^{-1}$ onwards in figure \ref{fig:Transition-Spacetime-MRI-wind-region}. This in turn causes a dramatic decrease of the Alfv\'en speed in both the lower and upper atmosphere due to the significant increase in density, and consequently the Alfv\'en point falls dramatically and becomes lower than the $n=2$ or $3$ MRI mode active region. The $n=2$ or $3$ mode which is responsible for driving the cycles is shut down, with the mid-plane bulge completely taking over the $b_x$ and $b_y$ profiles. The nature of the wind also changes to that of a slow wind launched from the disc by the magnetic pressure gradient of $b_y$.





\section{Investigation of the growing mid-plane bulge of the horizontal fields and its saturation}
\label{sec:bulge-sat}

\subsection{MRI linear stability analysis}
\label{sec:lin_stab_analysis}

To confirm our suspicion that the slowly but exponentially growing mid-plane horizontal fields are indeed a manifestation of the $n=1$ MRI mode, we perform a normal mode analysis on the equations looking for axisymmetric modes with frequency $\omega$. However, unlike the approach used in previous studies where simplifying assumptions are made about the background field variables, we allow for the background to take any values that form a valid disc solution, but not necessarily a steady state. The normal mode analysis solves the linearised equations, ignoring any time-dependence of the background state. The results should be meaningful if the background evolves sufficiently slowly, or perhaps if the variations (e.g. cyclic) in the background can be averaged over. 

We assume the standard ansatz
\begin{equation}
    \delta Q = \delta Q (z) \exp{(- i \omega t)}
\end{equation}
for the perturbations, where $Q$ denotes a generic field variable. The growth rate is then given by the imaginary part of $\omega$, which we label as $\sigma \equiv \imag{(\omega)}$. The full set of linearised equations are listed in Appendix \ref{app:linear_expansion}. Since the mid-plane bulge in $b_x$ and $b_y$ grows on a significantly longer timescale ($100$s of $\Omega^{-1}$) compared with the period of the wind cycles ($1-10$ $\Omega^{-1}$), we attempt to account for the growth rate of the bulge by computing the linear mode of each snapshot of the simulation as the background, and averaging the growth rate computed over the cycles to find the effective exponential growth rate that would be observed. Assuming that the mode growth has the form 
\begin{equation}
    f(t) \propto \exp{[ \gamma (t) ~ t]},
\end{equation}
where $\gamma (t)$ is the instantaneous growth rate at a particular point in time, then it can be shown that the effective growth rate over the time period from $t_1$ to $t_2$ would be given by (see derivation in Appendix \ref{app:average_growth_rate})
\begin{equation}\label{eq:B_eff_cal}
    \gamma_\text{eff} = \f{ \int_{t_1}^{t_2} \gamma (t) \rmd t}{t_2-t_1}.
\end{equation}

To solve the system of equations, we used a pseudo-spectral method with a decomposition on Whittaker cardinal functions (i.e. sinc functions) \citep{Boyd_2001}. The Whittaker functions naturally tend to $0$ as $\lvert z\rvert \to \infty$. { The equations were recast in terms of momenta $\delta \bmm \equiv \rho \delta \bmv$ instead of velocity $\delta \bmv$ to help with the convergence of solutions. In the case of the standard shearing box in the absence of an outflow, the magnetic fields are force-free in the low-density region at large $\lvert z\rvert$, and we would expect $\delta B_{x,y}$ to tend to $0$. $\rho$ would also tend to $0$ following the isothermal Gaussian profile as $\lvert z \rvert \to  \infty$, allowing us to use the momenta $\delta \bmm$ instead of velocity as suitable variables for the Whittaker basis. However, in our case, things are complicated not only by the presence of an outflow, but also the modified gravity. The density, $\rho$, no longer tends to $0$ as $\lvert z \rvert\to \infty$, and we should not expect either $\bmB$ or $\bmm$ to tend to $0$ at the boundaries. Nevertheless, as we are applying the solver to the simulation region of $\lvert z\rvert < 12H$, $\rho$ is still very small at the boundary and of the order $10^{-5}$. Although $\delta \bmB$ and $\delta \bmm$ do not technically vanish exponentially in the regimes we study, our solver was able to yield consistent results as the resolution was increased.} The full set of modified equations used in our pseudo-spectral method can be found in Appendix \ref{app:pseudospec}. As we have not applied any simplifying assumptions such as $v_z = 0$ or a pure vertical $B$ field, it is not possible to reduce the system of six equations (\ref{eq:MRI_lin1} to \ref{eq:MRI_lin6}) into any simpler form, as is usually done in other studies of the MRI linear modes \citep{SanoMiyama_1999,SalmeronWardle_2005}. We did, however, check that our solver yielded the same results as previous studies in these simplified regimes, as well as agreeing with the modes calculated in the vertical field only hydrostatic case using a simple shooting code solver, details of which can be found in Appendix \ref{app:shooting}. 

For most of our calculations, we used grids of $2201$ and $2301$ points over the domain $\lvert z \rvert < 12 H$ to analyse data from our $L_z=12 H$ simulations. Using two different resolutions allows us to assess the convergence of solutions, and also flags up cases for exclusion when modes are obscured by numerical oscillations which may be excited in the pseudo-spectral method by using specific grid resolutions. Generally, the modes showed good convergence, and no qualitative difference was observed from further increases in resolution. In order to speed up the calculations and also allow the matrices to be computationally soluble, we used the assumption that modes take either a slanted or hourglass symmetry about the mid-plane, reducing the number of elements in each dimension by half. This condition arises naturally when the background field variables also adopt a slanted or hourglass symmetry about the mid-plane, but its validity is more dubious in the case when the background is asymmetric. Our linear analysis is therefore better suited to studying the cyclic states, the steady state solution, and the early/late stages of the asymmetric transition between the hourglass symmetry cyclic and the slanted symmetry steady state where there is great semblance to one of the symmetries, while our results for the middle of the asymmetric transition period should be treated with caution. It should be noted though that the transition from slanted symmetry cyclic to slanted symmetry steady state always preserves the slanted symmetry, and there should not be any symmetry concerns regarding our method of calculation in that case.

\subsection{Direct analysis of the simulation states}

\subsubsection{Examining the cyclic states}
\label{sec:lin-cyclic}

As the mid-plane bulge begins its growth through amplifying perturbations of $b_{x}$ and $b_y$ in the cyclic states, we begin our investigation by applying our linear analysis directly to the simulations by using data of our $L_z=12 H$ hourglass symmetry cycle run \textbf{b1e5\_S} as the background disc. As we are primarily interested in whether the slow growing mid-plane mode can be understood through such analysis, we filtered out modes with an hourglass symmetry and only included the fastest growing mode with a slanted symmetry. However, it is interesting to note that the fastest growing modes obtained are almost exclusively of the slanted nature, with mode shapes resembling that of the growing mid-plane bulge observed. We attribute the lack of $n=2$ or $3$ modes to the rapidly changing dynamics of the cycles, which suggest that the modes are in the non-linear regime, and therefore not captured by our linear calculations. In order to remove modes obscured by rapid numerical oscillations, we compared results from two runs with different resolutions and only included modes whose growth rates differ by no more than $10\%$. We also applied a fast Fourier transform on the mode profiles and excluded modes dominated by extremely high frequencies. 

\begin{figure}
  \centering
    \includegraphics[width=0.5\textwidth]{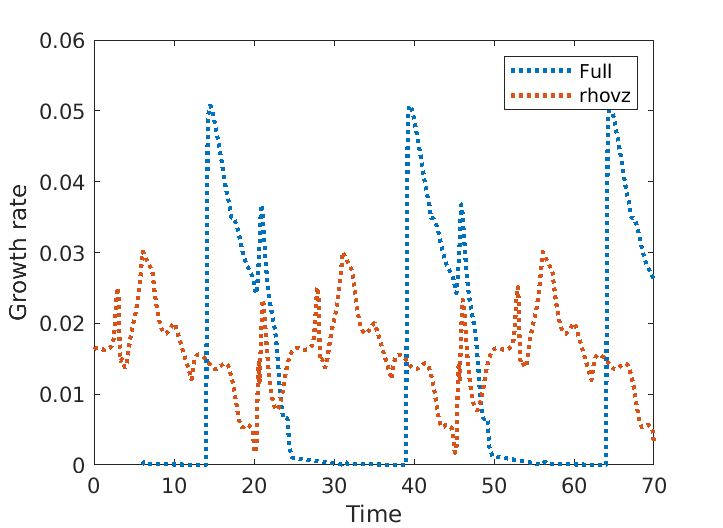}
    \caption{MRI growth rates obtained using linear analysis on the background fields of the hourglass symmetry cycles in run \textbf{b1e5\_S}. The blue dotted line is for full background, while the orange dotted line is when only $\rho$, $v_z$ and the background vertical field are included.}
    \label{fig:GR_time-bt10t5_bxSINmod}
\end{figure}

We did both calculations where we inputted all the variables from our simulations as the background state for the linear analysis, as well as ones where only $\rho, v_z$ and $B_z$ are included and the other variables $v_x$, $v_y$, $B_x$ and $B_y$ are set to $0$. The latter calculations, denoted with the suffix `rhovz', are motivated by the observation in the previous section of the importance of outflow and change in density profile in affecting the growth of the magnetic fields. Isolating these variables allows us to examine to what extent they are responsible for the behaviour we find. Fig. \ref{fig:GR_time-bt10t5_bxSINmod} shows the linearised growth rates calculated from the hourglass cycles in run \textbf{b1e5\_S} where $\beta_0=10^5$. The blue curve is for calculations with all variables, denoted `Full', while the orange curve is for the `rhovz' case. They both vary periodically with the phase of the cycles, although apart from the decrease in magnitude from $t=13~\Omega^{-1}$ up to the end of the peak at $t=22~\Omega^{-1}$, both the behaviour and magnitude of the growth rates are notably different. {\color{black}It is interesting to note that the times when the `Full' case yielded negligible growth rates for the slowly growing `bulge' mode are when the $b_{x}$ and $b_y$ peaks of the $n=2$ or $3$ mode in the background are beginning to significantly grow again in the MRI active region of $3.15 H < \lvert z \rvert < 4.75H$.} 

\begin{figure}
  \centering
    \includegraphics[width=0.5\textwidth]{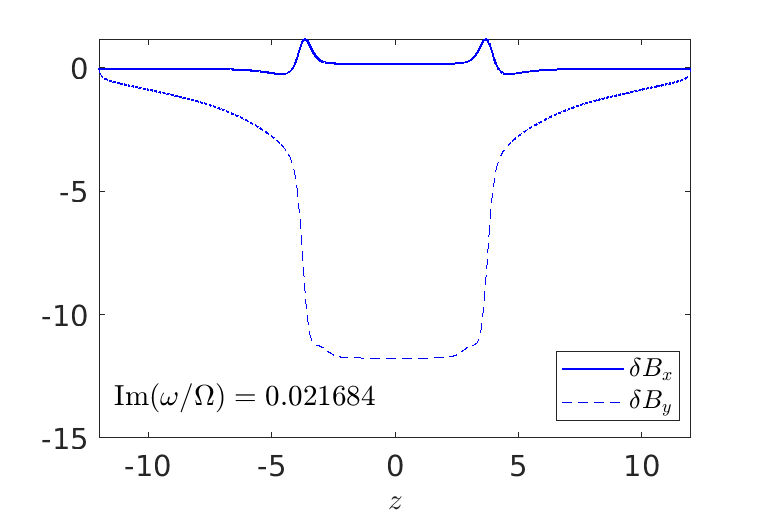}
    \caption{Mode at $t=22.2 \Omega^{-1}$ calculated from  \textbf{b1e5\_S} using the `Full' scheme.}
    \label{fig:mode-bt10t5_bxSINmod}
\end{figure}

The shape of the modes in both the `Full' and `rhovz' calculations highly resembles the horizontal magnetic fields of the slanted symmetry steady state profile reached at the end of the simulations. Fig. \ref{fig:mode-bt10t5_bxSINmod} shows the horizontal magnetic fields of one of the modes in the `Full' calculation at $t=22.2 \Omega^{-1}$. Like the slanted symmetry steady state, the profile has $\delta B_x$ peaks at $\lvert z\rvert \sim 3.15 H$, and a significantly larger magnitude $\delta B_y$ of the opposite sign in the mid-plane. We interpret the mode shape to be a result of the instability being active at $\lvert z\rvert \sim 3.15 H$ leading to the $b_x$ peaks there. The strong resistivity at the mid-plane causes the dense disc region to be linearly stable against the MRI, but the strong diffusion leaks the $b_x$ flux from the MRI active surface layers to the disc, resulting in a significant net $b_x$ in the mid-plane region. This is similar to the mechanism described in \citet{Turner_etal_2007}. Shearing of the $b_x$ field then generates the large $b_y$ that we see until saturation occurs.

When averaged using equation (\ref{eq:B_eff_cal}), the `Full' calculations give an effective linear growth rate of $\gamma_\text{eff} = 0.0122$, while the `rhovz' results give $\gamma_\text{eff} = 0.0155$. As we did not observe any mid-plane bulge growing in the run \textbf{b1e5\_S} (see Section \ref{sec:trans_description}), this growth rate should be
compared with the measured initial growth rate of $\sigma = 0.18$ in runs \textbf{b1e5\_A} and \textbf{b1e5\_G}. While the calculated growth rates are both slightly lower than the measured growths in the simulations, they are comparable and of the same order of magnitude. The difference in growth rates may be attributed to the effect of further changes in the background when the mid-plane bulge is already present and growing. Indeed, when we applied the linear mode solver using data from the runs \textbf{b1e5\_A} and \textbf{b1e5\_G} as background discs, we obtained closer values of $\gamma_\text{eff} = 0.0192 ~(\text{Full}), 0.0200 ~(\text{rhovz})$ and $\gamma_\text{eff} = 0.0185 ~(\text{Full}), 0.0143~ (\text{rhovz})$ respectively. This strongly suggests that the mid-plane bulge is indeed the result of a slow MRI mode of slanted symmetry growing on top of the cyclic state background.

\subsubsection{Robustness in behaviour across magnetisations}

We also repeated the same analysis for simulations with different $\beta_0$, the results of which are plotted in Figure \ref{fig:Line-beta0}. As can be seen, the calculated linear growth rate and its trend with $\beta_0$ closely follows that of the measured initial growth rates. This suggests that the mid-plane bulge growth is indeed a manifestation of the MRI across the different field strengths investigated.

\subsubsection{Progression to the slanted symmetry steady state}
\label{sec:saturation_process}

\begin{figure}
  \centering
    \includegraphics[width=0.5\textwidth]{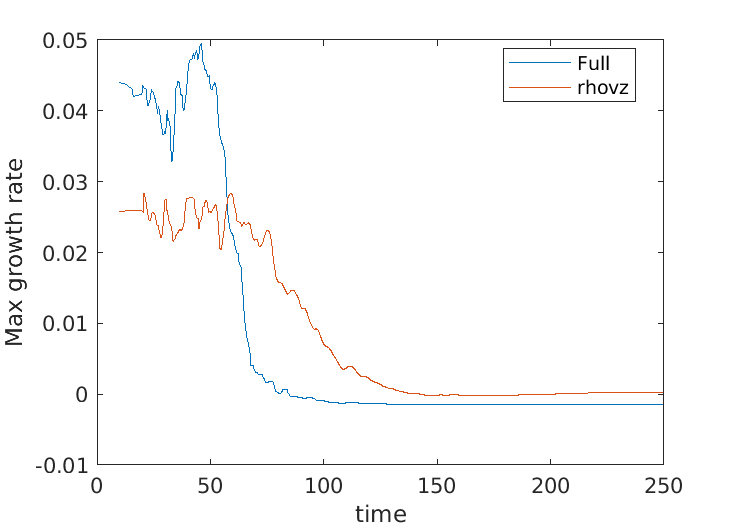}
    \includegraphics[width=0.5\textwidth]{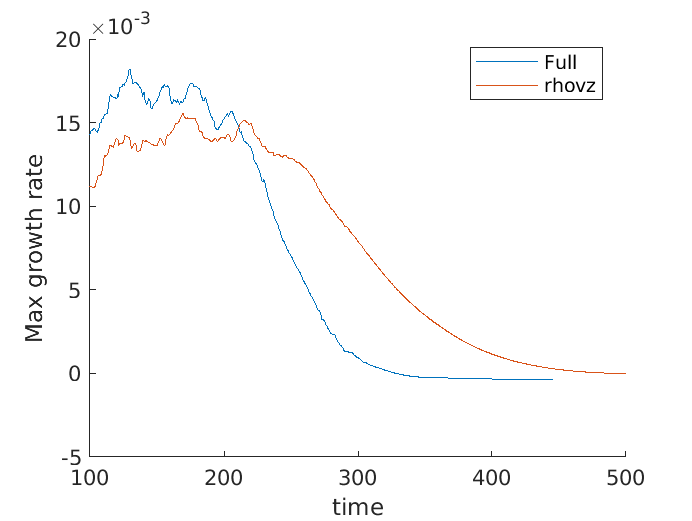}
    \caption{Plots of linear growth rate against time, calculated from \textbf{bt1e4\_G} (top) and \textbf{bt1e5\_G} (bottom).}
    \label{fig:Bgrowth-t-bt10t5_bxASYM}
\end{figure}

In order to investigate the saturation mechanism, we applied the linear growth rate calculations to our full disc runs  beginning in the cyclic stage (mostly hourglass), right up to them reaching the slanted symmetry steady states. Fig. \ref{fig:Bgrowth-t-bt10t5_bxASYM} shows a moving average of the growth rates (over $20 \Omega^{-1}$) calculated from \textbf{b1e4\_G} and \textbf{b1e5\_G}, with blue denoting the results for the `Full' scheme and orange for `rhovz'. We found that they all follow the same pattern. The linear growth rate varies in magnitude periodically with the cycles in the cyclic stage but with a moving average $\gamma_\text{eff}$ value close to the measured growth rate at those times. As the disc goes through the transition and the cycles shut down, the linear growth rate begins to gradually decrease in magnitude. During this time, the mid-plane bulge is also observed to slow down in its growth. Upon saturation to the slanted symmetry steady state, the largest linear growth rate becomes negative. While the `Full' and `rhovz' calculations have notable differences in the cyclic stage though yielding similar $\gamma_\text{eff}$ (see section \ref{sec:lin-cyclic}), their growth rates converge as the cycles shut down and the slanted symmetry steady state is reached. The rate of convergence is quickest for simulations with stronger magnetic fields ($\beta_0 \leq 10^4$). This suggests that for discs threaded by strong magnetic fields, the slowing of the growth of the mid-plane bulge and its eventual saturation are mostly due to the changes in the density profile and outflow, whereas in the case of weak magnetic fields, the mid-plane $b_y$ also has a significant effect in slowing the growth of the MRI. The eventual saturation however is still maintained by the density and outflow modifications, as shown by the convergence of the two curves as steady state is reached. 

\begin{figure}
  \centering
    \includegraphics[width=0.5\textwidth]{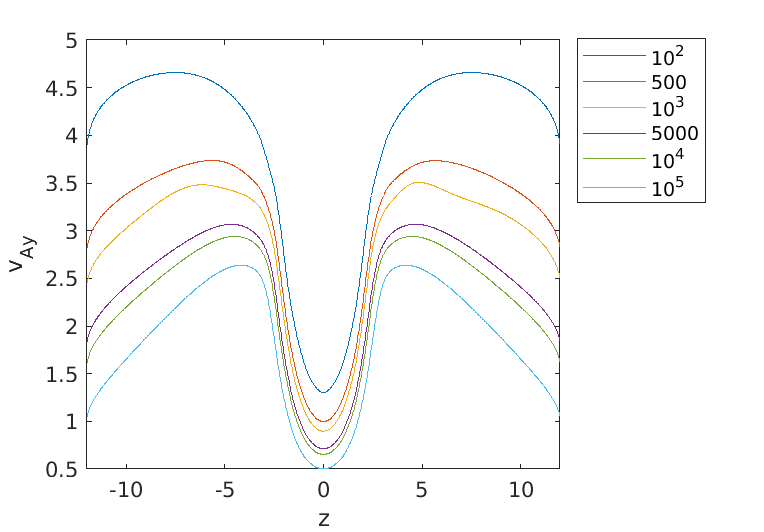}
    \caption{Vertical profiles of the azimuthal Alfv\'en speed (in units of the sound speed) for runs with different $\beta_0$, as indicated by the legend.}
    \label{fig:v_Ay-Steady-btVARY}
\end{figure}

\begin{figure}
  \centering
    \includegraphics[width=0.5\textwidth]{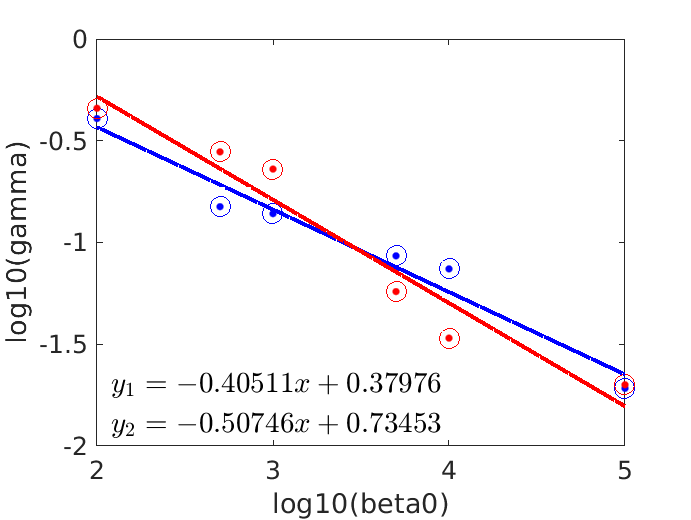}
    \caption{Plots of $\log$ of the time-averaged effective linear growth rate, $\gamma$, against $\log (\beta_0)$. Blue is for the case when all variables from the simulations are included as the background, while red is for when only $\rho$ and $v_z$ and $B_z$ from the simulations are included, and all other variables are set to $0$. The growth rates are calculated from full disc simulation runs. The blue and red lines, whose equations are given by $y_1$ and $y_2$ respectively, are the best fit lines for the `Full' and `rhovz' results. This plot is to be compared with Figure \ref{fig:sigma_v_beta0} of the growth rates observed in the simulations.}
    \label{fig:Line-beta0}
\end{figure}

The significance of a large $b_y$ in suppressing the mid-plane bulge growth rate was noted in the local dispersion analysis of the MRI by \citet{SanoMiyama_1999}. In section 3.2 (see also Figures 5 and 6) of their paper, they showed that the maximum growth rate is decreased as $B_\phi$ (equivalent to our $B_y$) is increased. They explained this effect by pointing out that the toroidal field acts as a magnetic pressure on the axisymmetric perturbations, suppressing the unstable growth. This effect was found to be significant when the azimuthal Alfv\'en speed, $v_{Ay}=\lvert B_y \rvert/\sqrt{\mu_0\rho}$, becomes faster than the sound speed. Figure \ref{fig:v_Ay-Steady-btVARY} plots the profiles of $v_{Ay}$ in the slanted symmetry steady states for our runs with different $\beta_0$. As is clearly shown, the $v_{Ay}$ values are of order unity when compared with the sound speed, with only a small decrease in magnitude as the magnetisation decreases, with the disc region $v_{Ay}$ becoming slightly lower than the sound speed at around $\beta_0=10^3$. This suggests that $B_y$ should indeed have a significant impact in suppressing MRI growth for the range of magnetisations considered. On the other hand, steady state runs with higher magnetisations (lower $\beta_0$) have higher mass outflows and likewise greater changes in their density profiles (to preserve mass conservation). We therefore hypothesise that the greater mass outflow and corresponding flattening of the density profile also has an effect of suppressing the mid-plane MRI growth, and is the dominant mechanism for saturation in the low $\beta_0$ cases. 


\subsection{The significance of density modification, outflow and azimuthal magnetic field on the MRI}

To assess the importance of outflow, density profile modification, and the azimuthal field in suppressing the $n=1$ mode of the MRI, we used a simplified model where we ignored the physics of wind-launching, and the backgrounds are comprised of analytic profiles governed by easy to interpret parameters. For the density, we used a profile of the form
\begin{equation}
    \rho = \rho_0 \exp{(-z^2/2)} 
    + \rho_1 \left( \f{\lvert z \rvert}{1+z^2} \right),
\end{equation}
where $\rho_0$ and $\rho_1$ are parameters governing the normalisation of the density and the flattening out of the hydrostatic Gaussian density profile due to an outflow respectively. We keep $\rho_0=1$ for all our runs so that as before the magnetisation of the disc is set through varying $B_z$, while the larger the value of $\rho_1$, the more flattened the profile becomes.
The vertical velocity is calculated through
\begin{equation}
    \lvert \rho v_z \rvert = \Dot{m}_w = \cst ,
\end{equation}
where $\Dot{m}_w$ is the outflow rate and is one of our input parameters.
It can then be shown in our model that, for $\lvert z \rvert \gg 1$, 
\begin{equation}
    \rho \sim \f{\rho_1}{\lvert z \rvert},
\end{equation}
and
\begin{equation}
    v_z \sim \f{\Dot{m}_w}{\rho_1} z .
\end{equation}
The vertical magnetic field strength $B_z$ is set by the $\beta_0$ parameters as before, and we used the same resistivity profile with $\eta_0=2$ and $\eta_\infty=0.01$ as in most of our runs. For the azimuthal field, we set the value of $B_y$ such that $v_{Ay}$, which we use as the input parameter, is always constant, and is given by
\begin{equation}
    B_y = \sqrt{v_{Ay} \rho}.
\end{equation}
All other field variables are set to $0$, and we ignore any additional physics that might be operating in the disc. After the linear calculation, in most cases we extract the $n=1$ mode by restricting our result to the highest growth rate mode of the slanted symmetry, which we also checked to corresponds to the expected form of a mode with no nodes. When we reduced $\Dot{m}_w$ to $0$ however, we found that for high $\beta_0$ discs, the highest growth rate mode of slanted symmetry is no longer the $n=1$ mode as before. This is in agreement with the calculations of \citet{Latter_etal_2010}, who showed that as disc magnetisation is weakened, the highest growth rate mode moves from the $n=1$ mode to higher order modes. However, for the background disc parameters relevant to our study of the saturation to the slanted wind state (particularly the condition that $z_{Az} < 4H$), the $n=1$ mode is always the fastest growing mode. 

Our results are plotted in Fig. \ref{fig:LinSimp} in the following manner. Each dot corresponds to a result from a different background profile, with the $\log_{10}$ of the linear growth rate calculated shown by its colour. Each vertical column of dots has the same set of parameters with the exception of the mass outflow, which is indicated by the vertical coordinate of the Alfv\'en point of the background profile, with a lower Alfv\'en point corresponding to a larger mass outflow. The vertical columns are grouped horizontally in clusters according to their $\beta_0$ values of $10^2, 10^3, 10^4$ and $10^5$. The top plot has clusters of three columns for each $\beta_0$ value, where $\rho_1$ varies in the order $0.0005$, $0.005, 0.05$ from left to right, with $v_{Ay}$ kept at $0$ for all these runs. The bottom plot has clusters of five columns for each $\beta_0$ value, where $v_{Ay}$ varies in the order $0,0.1,0.5,0.8,1$ from left to right, while $\rho_1$ is always $0.0005$ for these runs. Figure \ref{fig:LinSimp_vz0} plots the variation of the background azimuthal field, characterised by $v_{Ay}$, against the magnetisation, characterised by $\beta_0$. The growth rate of the fastest growing mode is indicted by its colour, but is not necessarily the $n=1$ mode, as discussed above. The columns in the triplet for each $\beta_0$ value correspond from left to right to the three density modifications of $\rho_1=0.0005, 0.005$ and $0.05$, while $\Dot{m}_w$ is set to $0$ indicating no outflow for these calculations.


\subsubsection{Effect of density modification}

We begin by examining the effect of modifying the density distribution. As the density profile becomes flatter with a larger $\rho_1$ (moving across the columns within each triplet in the top plot of Figure \ref{fig:LinSimp}), the linear growth rate of the $n=1$ mode decreases for almost all outflow strengths and disc magnetisations explored. The only exception is the case when there is a large $v_{Ay}$ and no outflow present. We attribute the general trend to the decrease in the Ohmic Elsasser number $\Lambda$ in the atmospheric regions as the density profile becomes flatter, since $\Lambda \propto 1/\rho$. This is turn extends the region over which the quenching effect of Ohmic resistivity on the MRI is significant, lowering the overall growth rate. We can see this reflected in the mode shapes as $\delta B_x$ and $\delta B_y$ vanish at the boundaries less rapidly from the central bulge as $\rho_1$ is increased. Generally, the effect of density modification corresponds to a 10-fold decrease in the linear growth rate when the density in the atmosphere is increased 100-fold. This most likely is a significant contributing factor to the shutdown of the $n=1$ MRI mode in the slanted wind, as the atmospheric density does experience a roughly 100-fold increase in the simulations compared with that of the cyclic state at all the magnetisations explored. However, Figure \ref{fig:LinSimp_vz0} shows us that in the absence of outflow and an azimuthal field, a significant linear growth rate of $\mathcal{O}(0.1)$ still remains, so density modification alone is not sufficient to account for the saturation of the steady wind.

\subsubsection{Effect of outflow}

The presence of an outflow drastically reduces the linear growth rate when the Alfv\'en point $z_{Az}$ is lowered beyond $\lvert z \rvert \sim 3.15H$. This corresponds to a strong and dense wind launched from the lower atmosphere below the $\delta B_x$ peaks of the $n=1$ mode, which we suggested back in section \ref{sec:lin-cyclic} when coupled with diffusion of the horizontal field to the mid-plane may be responsible for driving the mode development. Hence, once $z_{Az}$ is lowered below the peaks, we expect vertical outward advection of the MRI mode to dominate, shutting down the MRI completely, which is indeed what we find. The behaviour of the MRI when a weaker wind is present such that $z_{Az} > 4H$ in our background configuration is less clear, and there are indications that under certain conditions, such as when $\beta_0=10^3$ and $\rho_1 = 0.005$ (2nd column of the 2nd cluster from the left of the top plot of Figure \ref{fig:LinSimp}), as the wind weakens, there may be a brief shutdown of the $n=1$ mode before its growth rate is restored to its high no-outflow value (Figure \ref{fig:LinSimp_vz0}). However, as the Alfv\'en points in the saturation phase of our simulations is always below $\lvert z \rvert=4H$, we can conclude that a strong outflow ($z_{Az} < 3.15H$) induced advection of the $n=1$ mode does have a critical effect of its eventual shutdown. The rapid decrease of the growth rate to $0$ as $z_{Az}<3.15H$ is also most prominent for more highly magnetised (lower $\beta_0$) discs, a result which is in line with the generally higher Alfv\'en points we found for the saturated slanted winds for lower $\beta_0$ discs. 

\subsubsection{Effect of azimuthal field}
\label{sec:lin_B_y_model}

\begin{figure}
  \centering
    \includegraphics[width=0.5\textwidth]{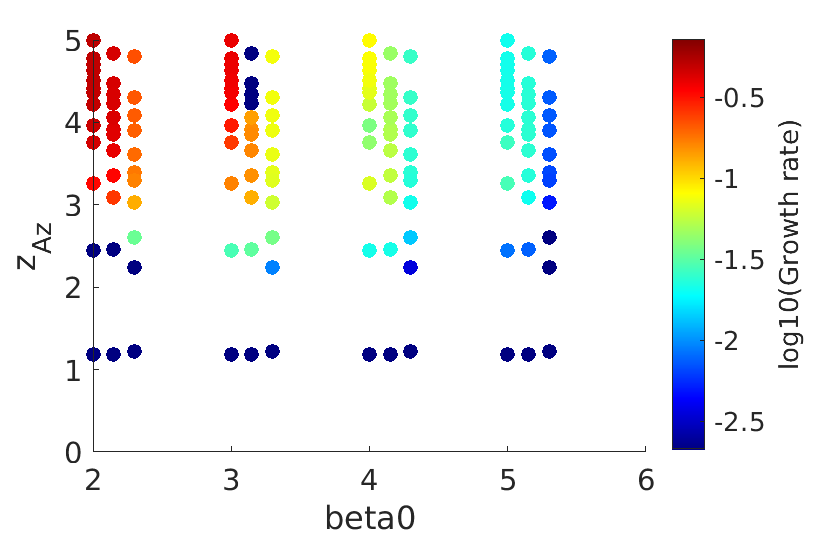}
    \includegraphics[width=0.5\textwidth]{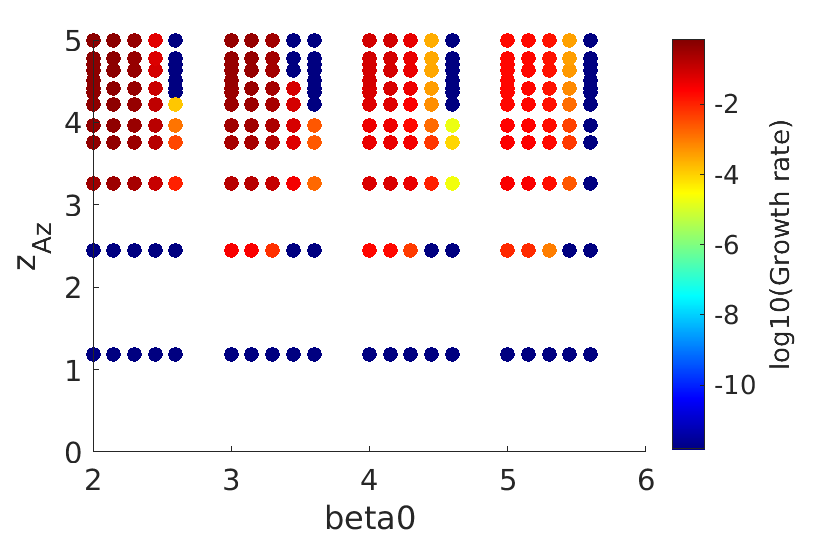}
    \caption{Plots of log10 of the linear growth rate with $\beta_0$ of the background profile varying horizontally taking the values $10^2,10^3,10^4,10^5$, and $z_{Az}$ varying vertically. The top plot has $v_{Ay}=0$ for all data points, while each triplet of columns with the same $\beta_0$ value has $\rho_1=0.0005,0.005$ and $0.05$ from left to right. The bottom plot fixes $\rho_1=0.0005$ for all data points, while each quintet of columns with the same $\beta_0$ value has $v_{Ay}=0,0.1,0.5,0.8,1$ from left to right.}
    \label{fig:LinSimp}
\end{figure}
\begin{figure}
  \centering
    \includegraphics[width=0.5\textwidth]{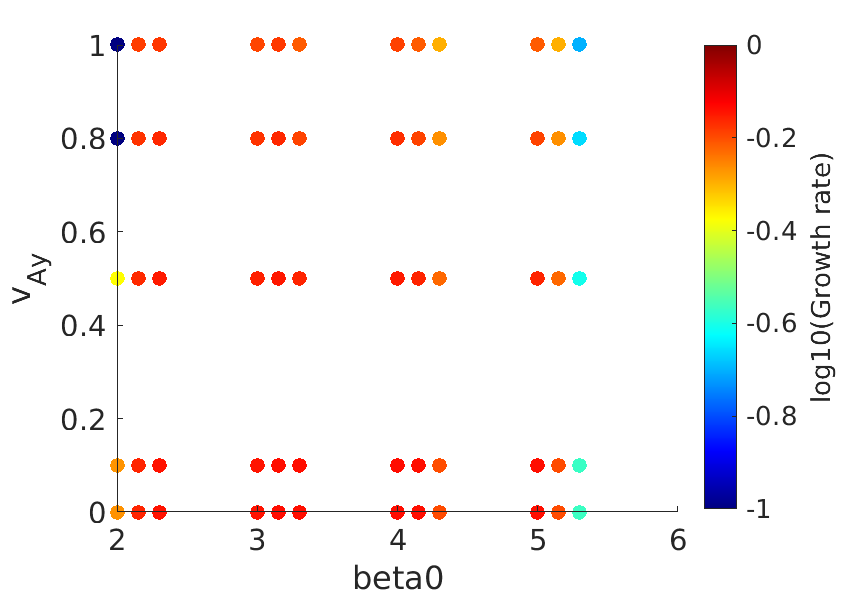}
    \caption{Plot of log10 of the fastest linear growth rate (not necessarily the $n=1$ mode) with $\beta_0$ of the background profile varying horizontally taking the values $10^2,10^3,10^4,10^5$, and $v_{Ay}$ varying vertically. Each triplet of columns with the same $\beta_0$ value has $\rho_1=0.0005,0.005$ and $0.05$ from left to right. }
    \label{fig:LinSimp_vz0}
\end{figure}

Finally, we examine the effect of azimuthal field strength. As $v_{Ay}$ increases from $0$ to $1$ (left to right in each cluster of five columns in the bottom plot of Figure \ref{fig:LinSimp}), in the majority of cases, the growth rate is abruptly and rapidly reduced to $0$ as $v_{Ay}$ increases beyond $0.8$. There is a limited range of intermediate outflow strengths (where $3H < z_{Az} <4H$) when the quenching effect of the azimuthal field on the MRI is less significant, particularly for the more highly magnetised discs. However, for the parameters that most closely resemble the saturated states of our simulations, the effect of $v_{Ay} \sim 1$ is indeed significant in contributing to the shutdown of the MRI. This, coupled with our analysis in Section \ref{sec:saturation_process}, suggests that the saturation mechanism is largely a combination of vertical advective damping from the outflow, and a large azimuthal field strength in quenching axisymmetric perturbations. 

\section{Discussion and astrophysical implications}
\label{sec:discussion}

\subsection{Summary of the results}

By using radially local 1D vertical resistive shearing box simulations in the parameter regime relevant to protoplanetary discs, we have found wind solutions which go through three stages of development: cyclic, transitive and slanted symmetry steady winds, the last of which bear great resemblance to the slanted winds seen in other local and global simulations. We have assessed, in particular, the importance of large-scale MRI channel modes in driving these wind states. Figure \ref{fig:CyclesScheme} shows the mechanism we proposed to be responsible for the cyclic state, which is driven by periodic excitation of the $n=2$ or $3$ MRI channel mode, coupled with advective eviction when the Alfven point falls below the mode peaks. We have shown that the mid-plane bulge which eventually causes the transition to the steady wind is a result of a much slower growing $n=1$ MRI mode, and the transition occurs when the mid-plane $B_y$ value is larger than the maximum peak $B_y$ strength in the cyclic stage. Saturation of the growing bulge to the steady state wind of slanted symmetry occurs due to both a combination of advective damping from the strong wind, and suppression of the instability from a large toroidal field. We also found that a more magnetised disc would speed up the process of transition and saturation to the steady wind through our parameter study, and confirmed the robustness of our results by varying both the box size and mass replenishment schemes.

\subsection{Connection of our results with other MRI-driven wind simulations}

The first implication of our results is in understanding the essential ingredients for the development of the slanted wind observed in both local \citep{Bai_2013,Lesur_etal2014} and global simulations \citep{Bethuneetal2017,Bai_2017,Rodenkirch_etal_2020,Gressel_etal_2020,Riols_etal_2020}. We have shown that it is the result of a slowly growing $n=1$ MRI mode, characterised by a mid-plane bulge in $B_x$ and $B_y$. This bulge eventually flips the disc symmetry from the traditional hourglass configuration to the slanted wind, shutting down other MRI modes that may be present via advective damping from the dense low-Alfv\'en point winds launched. This process seems to only require the presence of a strongly diffusive mid-plane region, which forces all higher order modes than the $n=1$ one to be localised in the disc lower atmosphere or above. These modes are in turn are shut down as the Alfv\'en point falls below those regions, allowing the slower $n=1$ mode to grow and eventually dominate the profile. Therefore, we should expect all discs with highly Ohmic diffusive mid-plane regions to eventually settle into the slanted state regardless of its history, which is in line with what has been reported by certain authors in global simulations \citep{Bethuneetal2017,Riols_etal_2020}. In particular, the simulations of \citet{Bethuneetal2017} showed a convergence to the slanted profile on a timescale of 100s of $\Omega^{-1}$ (see their Figure 23), which is in agreement with the timescales we have found in our study. Our parameter study suggests that we should expect lower $\beta_0$ discs (i.e. discs with higher magnetisations) to be more prone and quicker to develop the slanted symmetry wind state. This is also noted in other local simulations \citep{BaiStone2013}, while figure 31 of \citet{Bethuneetal2017} also appears to indicate the same trend, but it would be interesting for future global simulations to explore this in more detail as the parameter space is further expanded and simulations are run for longer in the future.

The periodic outbursts observed in our cyclic stage are reminiscent of cyclic outbursts found in other local simulations \citep{SuzukiInutsuka_2009,Suzuki_etal_2010,Fromangetal2013,Riols_etal_2016}. Similarly to our cycles, their simulations show a strong correlation between development of the horizontal magnetic fields and the outburst behaviour, suggesting that horizontal magnetic fields are invovled in the cyclic launching mechanism. As we have already discussed in Section \ref{sec:RiolsComp}, our outbursts are different in nature and mechanism from those of \citet{Riols_etal_2016}, which is most likely due to our lower field strengths and the dead-zone resistivity profile we used. There are greater similarities in both the outburst strength, period and launch region between our results and those of \citet{Suzuki_etal_2010}, which also used a dead-zone profile over and $\beta_0=10^6$, but over a much smaller box size ($L_z=4H$). Their simulations were done in 3D, and it was noted that their disc winds were partly driven by the breakup of channel flows triggered by the MRI in the lower atmosphere, although they did not do a detailed analysis of the mechanism as we have done here. This confirms that even though are our simulations are in 1D, they are nevertheless able to capture one of the key behaviours that may be responsible for driving periodic outbursts. We are also able to conduct simulations for longer and with lower $\beta_0$ values than \citet{Suzuki_etal_2010}, hence showing that the slanted symmetry state would eventually take over and change the wind behaviour, which was hinted at in their snapshots by the slightly slanted fields at the mid-plane, but were not fully developed to the extent that the cycles would be shut down and morph into the slanted steady wind, due to both their shorter run time and the very high $\beta_0$ value they used. Outburst behaviour was also observed in the initial stages of the Ohmic-only global simulations of \citet{Rodenkirch_etal_2020} before settling of the disc to the slanted symmetry wind state, and it would be interesting to investigate to what extent the cyclic outburst behaviour we uncovered is also present in global discs.

It is also worth comparing our 1D resistive shearing box calculations with simulations where other non-ideal MHD effects are present. Notably, the simulations of \citet{BaiStone2013} showed no periodic wind solutions despite similar parameter regimes to us in resistivity and field strength ($\beta_0=10^6-10^3$). We attribute this difference to the presence of ambipolar diffusion in their lower atmosphere (absent in ours), which may have stemmed the growth of the $n=2$ or $3$ MRI mode peaks in $b_x$ and $b_y$ before they are strong enough to modify the density profile sufficiently to cause the occurrence of a second Alfv\'en point beneath the peaks to drive the outbursts. The slanted wind profiles in \citet{BaiStone2013} also have a longer and quasi-steady transition state, where a strong current layer is maintained at $z\sim 3H$ for about 100 orbits, before a full slanted wind solution is recovered. Again, we attribute this difference to the presence of ambipolar diffusion in their simulations, which would have altered the shapes and growth rates of the MRI channel modes. Simulations where the Hall effect is also included \citep{Lesur_etal2014,Bai_2014,Bai_2015,Simon_etal2015} showed that its presence may enhance the development of a mid-plane azimuthal magnetic field and progression to the slanted wind state, depending on which polarity the Hall term has with respect to the vertical field. All this shows that the additional of other non-ideal effects presents a wide parameter space for exploration which could have significant enhancements and changes to our Ohmic only picture, and will be investigated in a future paper.

\subsection{Implications on our understanding of protoplanetary disc dynamics}

One area of potential interest is in how the transition of the disc to the slanted wind state impacts on the accretion and radial transport of $B_z$ flux in the disc. By nature of the slanted symmetry, a disc with such configuration cannot (at least in the local model) support a net radial steady state transport of matter or $B_z$ flux, as contributions from both sides cancel out. Global simulations \citep{Bai_2017,Bethuneetal2017,Gressel_etal_2020,Riols_etal_2020} have also shown that such symmetry may lead to a reduction in both overall accretion and flux transport rate, and may even cause the disc wind and accretion stream to be restricted to one hemisphere only. Since the slanted symmetry steady state is more easily reached when the local $B_z$ flux is strong, it may contribute to an automatic shut-down mechanism for the flux transport when the local build-up of $B_z$ flux becomes too strong and the disc transitions to the slanted state. This in turn, could have an interplay with the magnetic wind driven ring formation mechanism recently uncovered by \citet{RiolsLesur_2019,Riols_etal_2020}, which assumes the wind to already have the slanted symmetry in the more highly magnetised gap regions. A future study probing the importance of the transition to the slanted wind state for the working of this mechanism, as well as the long term radial transport of vertical flux would be needed to address these questions. 

Finally, the periodic outbursts observed in our cyclic stage show that MRI-wind outburst cycles could in theory operate in the PPD regime, and may be linked to the time variability observed in some PPDs \citep{Wisniewski_etal_2008,Muzerolle_etal_2009,Bary_etal_2009}. However, the simplified nature of our study does not allow us to form any firm conclusion on the possible link between our cycles and observations. Future work will need to be done with more realistic disc profiles, as well as addressing the problem in the global simulations, to ascertain if such connections exist.

\section*{Acknowledgements}

The authors would like to thank the annonymous reviewer for a prompt and detailed report on the manuscript. PKCL would like to thank the Croucher Foundation and the Cambridge Commonwealth, European \& International Trust for their generous support in funding his PhD studentship through a Cambridge Croucher International Scholarship.

\section*{Data availability}

Data used in this paper is available from the authors upon reasonable request.




\bibliographystyle{mnras}
\bibliography{references} 



\appendix

\section{Linear expansion of the equations on a general background}
\label{app:linear_expansion}

We study the behaviour of the MRI modes using a linear expansion of the perturbations on top of a general background assumed to vary on a longer timescale.
Using the ansatz 
\begin{equation}
    \delta Q = \delta Q (z) \exp{ (- i \omega t)}
\end{equation}
for perturbations $\delta Q$, and using the notation $D \equiv \p / \p  z$, the linearised equations are 
\begin{equation}\label{eq:MRI_lin1}
\begin{aligned}
    - \rho v_z D \delta v_x & + B_z D \delta B_x  \\
    & = B_z (D B_x) \f{\delta \rho}{\rho}
    - i \omega \rho \delta v_x 
    - 2 \rho \Omega \delta v_y
    + \rho (D v_x) \delta v_z , 
\end{aligned}
\end{equation}
\begin{equation}
\begin{aligned}
    - \rho v_z D \delta v_y & + B_z D \delta B_y \\
    & = 
    B_z (D B_y) \f{\delta \rho}{\rho}
    + \f{1}{2}\rho \Omega \delta v_x
    -i \omega \rho \delta v_y
    + \rho (D v_y) \delta v_z ,
\end{aligned}
\end{equation}
\begin{equation}
\begin{aligned}
    D \delta p
    + \rho v_z D & \delta v_z 
    + B_x D \delta B_x
    + B_y D \delta B_y \\
    = & ~ (D p + B_y D B_y + B_x D B_x) \f{\delta \rho}{\rho}
    + \rho (i \omega - D v_z )\delta v_z \\
    & - (D B_x) \delta B_x
    - (D B_y) \delta B_y ,
\end{aligned}
\end{equation}
\begin{equation}
    v_z D \delta \rho
    + \rho D \delta v_z
    = (i\omega - D v_z)\delta \rho 
    - (D \rho)\delta v_z,
\end{equation}
\begin{equation}
\begin{aligned}
    \eta D^2 \delta B_x
    + B_z D & \delta v_x
    - B_x D \delta v_z
    - (v_z - D \eta) D \delta B_x \\
    = & ~ 
     (D B_x)\delta v_z 
    - (i \omega - D v_z)\delta B_x,
\end{aligned}
\end{equation}
\begin{equation}\label{eq:MRI_lin6}
\begin{aligned}
    - \eta D^2 \delta B_y
    - B_z D & \delta v_y
    + B_y D \delta v_z
    + (v_z - D \eta) D \delta B_y \\
    = & ~
    - (D B_y)\delta v_z
    - \f{3}{2} \Omega \delta B_x
    + (i \omega - D v_z)\delta B_y.
\end{aligned}
\end{equation}
We have six equations and six unknowns:
\begin{equation}
    \delta \rho, \delta v_x, \delta v_y, \delta v_z, 
    \delta B_x, \delta B_y,
\end{equation}
and one eigenvalue,
\begin{equation}
    \omega.
\end{equation}
This is therefore a complete system of equations for obtaining a solution in combination with boundary conditions and an arbitrary normalisation condition.

\subsection{Recasting into operator form for pseudospectral analysis}
\label{app:pseudospec}

As described in section \ref{sec:lin_stab_analysis}, it is more useful to use $m_{x,y,z}$ than $\delta v_{x,y,z}$ to encourage convergence when using Whittaker functions. The corresponding changes are given by
\begin{equation}
    \delta v_{x,y,z} \to \f{1}{\rho} \delta m_{x,y,z}.
\end{equation}

Recasting the equations in operator form (using isothermal equation of state with $c_s^2=1$):
\begin{equation}
\begin{aligned}
    - B_z (D & B_x) \f{\delta \rho}{\rho}
    + \left[ v_z \left( \f{D \rho}{\rho}  - D \right)
    + i \omega
    \right]\delta m_x \\
    &
    + 2 \Omega \delta m_y
    - (D v_x) \delta m_z
    + B_z D \delta B_x  
    = 0 , 
\end{aligned}
\end{equation}
\begin{equation}
\begin{aligned}
    - B_z (D & B_y) \f{\delta \rho}{\rho}
    - \f{1}{2} \Omega \delta m_x \\
    &
    + \left[ v_z \left( \f{D \rho}{\rho} - D \right) 
    + i \omega \right]
    \delta m_y
    - (D v_y) \delta m_z
    + B_z D \delta B_y = 0 ,
\end{aligned}
\end{equation}
\begin{equation}
\begin{aligned}
    & \left[ D - \f{1}{\rho} 
    (D \rho + B_y D B_y + B_x D B_x )
    \right] \delta \rho\\
    & \qquad + \left[ v_z \left( D - \f{D \rho}{\rho} \right)
    - i \omega + D v_z \right]
    \delta m_z \\
    & \qquad
    + ( B_x D + D B_x)\delta B_x
    + ( B_y D + D B_y)\delta B_y
    = 0,
\end{aligned}
\end{equation}
\begin{equation}
    ( v_z D 
    - i\omega + D v_z
    )\delta \rho
    + D \delta m_z
    =  0,
\end{equation}
\begin{equation}
\begin{aligned}
    \eta \rho D^2 & \delta B_x 
    + B_z \left( D - \f{D \rho }{\rho} \right) \delta m_x \\
    & + \left[ B_x \left( - D + \f{D \rho }{\rho} \right)
    - D B_x \right]
    \delta m_z \\
    & 
    + \left[ (- v_z + D \eta) \rho D 
    + \rho (i \omega - D v_z) \right]
    \delta B_x
    = 
    0,
\end{aligned}
\end{equation}
\begin{equation}
\begin{aligned}
    - \eta \rho D^2 & \delta B_y 
    + B_z \left( - D + \f{D \rho }{\rho} \right) \delta m_y \\
    & + \left[ B_y \left( D - \f{D \rho }{\rho} 
    \right) 
    + D B_y
    \right] \delta m_z \\
    & 
    + \f{3}{2} \rho \Omega \delta B_x
    + [(v_z - D \eta) \rho D 
    + (- i \omega + D v_z)\rho 
    ]\delta B_y
    = 0.
\end{aligned}
\end{equation}

\subsection{Simplied regime: reduced equations and the shooting method}
\label{app:shooting}

In the limit, $v_z, B_x, B_y, \delta v_z = 0$, we obtain:
\begin{equation}
    B_z D \delta B_x  
    = 
    - i \omega \rho \delta v_x 
    - 2 \rho \Omega \delta v_y , 
\end{equation}
\begin{equation}
    B_z D \delta B_y = 
    \f{1}{2}\rho \Omega \delta v_x
    -i \omega \rho \delta v_y ,
\end{equation}
\begin{equation}
\begin{aligned}
    0
    = (D p)  \f{\delta \rho}{\rho}
    - D \delta p,
\end{aligned}
\end{equation}
\begin{equation}
    0
    = i\omega \delta \rho,
\end{equation}
\begin{equation}
\begin{aligned}
    \eta D^2 \delta B_x
    + B_z D \delta v_x
    + ( D \eta) D \delta B_x
    =  
    - i \omega \delta B_x,
\end{aligned}
\end{equation}
\begin{equation}
\begin{aligned}
    - \eta D^2 \delta B_y
    - B_z D \delta v_y
    - ( D \eta) D \delta B_y
    = & ~
    - \f{3}{2} \Omega \delta B_x
    + i \omega \delta B_y.
\end{aligned}
\end{equation}
Notice that if a mode exists, $\delta \rho = 0$, hence we only have four equations effectively.

Using an isothermal equation of state, and units such as $c_s^2 = 1$, this becomes
\begin{equation}
    D \delta B_x  
    = 
    - i \omega \f{\rho}{B_z} \delta v_x 
    - 2 \f{\rho}{B_z} \Omega \delta v_y , 
\end{equation}
\begin{equation}
    D \delta B_y = 
    \f{1}{2}\f{\rho}{B_z} \Omega \delta v_x
    -i \omega \f{\rho}{B_z} \delta v_y ,
\end{equation}
\begin{equation}
\begin{aligned}
    \eta D^2 \delta B_x
    + B_z D \delta v_x
    + ( D \eta) D \delta B_x
    =  
    - i \omega \delta B_x,
\end{aligned}
\end{equation}
\begin{equation}
\begin{aligned}
    \eta D^2 \delta B_y
    + B_z D \delta v_y
    + ( D \eta) D \delta B_y
    = & ~
    \f{3}{2} \Omega \delta B_x
    - i \omega \delta B_y.
\end{aligned}
\end{equation}

We can further express $\delta v_x$ and $\delta v_y$ in terms of $D \delta B_x$ and $D \delta B_y$:
\begin{equation}
    \delta v_x = \f{1}{\Omega^2 - \omega^2}
    \left(\f{B_z}{\rho}\right)
    \left( - i \omega D \delta B_x
    + 2 \Omega D \delta B_y \right),
\end{equation}
\begin{equation}
    \delta v_y = -\f{1}{\Omega^2 - \omega^2}
    \left(\f{B_z}{\rho}\right)
    \left( \f{\Omega}{2} D \delta B_x
    + i \omega D \delta B_y \right).
\end{equation}

Substituting this into the second order equations, we get:
\begin{equation}
\begin{aligned}
    \eta D^2 \delta B_x
    & + \f{v_{Az}^2}{\Omega^2 - \omega^2} 
    \Bigg( - i \omega D^2 \delta B_x
    + 2 \Omega D^2 \delta B_y \\ &
    \qquad \qquad ~ ~ ~ ~ ~
    - \f{D \rho}{\rho} \left[ 
    - i \omega D \delta B_x
    + 2 \Omega D \delta B_y \right]
    \Bigg) \\
    & 
    + ( D \eta) D \delta B_x
    =  
    - i \omega \delta B_x,
\end{aligned}
\end{equation}
\begin{equation}
\begin{aligned}
    \eta D^2 \delta B_y &
    - \f{v_{Az}^2}{\Omega^2 - \omega^2} 
    \Bigg( \f{\Omega}{2} D^2 \delta B_x
    + i \omega D^2 \delta B_y \\
    & \qquad \qquad ~ ~ ~ ~
    - \f{D \rho}{\rho} \left[ 
    \f{\Omega}{2} D \delta B_x
    + i \omega D \delta B_y
     \right]
    \Bigg) \\ &
    + ( D \eta) D \delta B_y
    = 
    \f{3}{2} \Omega \delta B_x
    - i \omega \delta B_y.
\end{aligned}
\end{equation}
Using the same definitions as Sano \& Miyama, where
\begin{equation}
    \sigma = (\Omega^2 - \omega^2 ) \f{\eta }{v_{Az}^2} - i \omega,
\end{equation}
\begin{equation}
\begin{aligned}
    \mathcal{S} 
    = & \left[  
    i \omega \f{D \rho}{\rho} 
    + \f{(\Omega^2 - \omega^2)}{v_{Az}^2} D \eta \right] D \delta B_x 
    \\ &
    - 2 \Omega \f{D \rho}{\rho} D \delta B_y
    + i \omega \f{(\Omega^2 - \omega^2)}{v_{Az}^2}\delta B_x,
\end{aligned}
\end{equation}
\begin{equation}
\begin{aligned}
    \mathcal{T}
    = & \f{\Omega}{2} \f{D \rho}{\rho} D \delta B_x
    + \left[ i \omega \f{D \rho}{\rho}
    + \f{(\Omega^2 - \omega^2)}{v_{Az}^2} D \eta
    \right] D \delta B_y \\ &
    - \f{(\Omega^2 - \omega^2)}{v_{Az}^2}
    \left [ \f{3}{2} \Omega \delta B_x
    + i \omega \delta B_y \right],
\end{aligned}
\end{equation}
the equations are recast as
\begin{equation}
    \sigma D^2 \delta B_x + 2 \Omega D^2 \delta B_y
    = - \mathcal{S},
\end{equation}
\begin{equation}
    - \f{\Omega}{2} D^2 \delta B_x + \sigma D^2 \delta B_y
    = - \mathcal{T}.
\end{equation}
This gives us finally
\begin{equation}
    D^2 \delta B_x = \f{- \sigma \mathcal{S} + 2 \Omega \mathcal{\tau}}
    {\Omega^2 + \sigma^2},
\end{equation}
\begin{equation}
    D^2 \delta B_y = \f{-(\Omega/2) \mathcal{S} - \sigma \mathcal{\tau}}
    {\Omega^2 + \sigma^2},
\end{equation}
which reproduces the results of \citet{SanoMiyama_1999}.

\section{Calculating the average exponential growth}
\label{app:average_growth_rate}

The linear MRI analysis yielded a varying growth rate, the average effect of which may be able to explain the slow exponential increase we observe in the mid-plane $b_{x}$ and $b_y$ values. We need a suitable average to see its effective exponential growth over one cycle.

In the constant exponential growth rate model:
\begin{equation}
    f(t) = A \exp{(\gamma t)},
\end{equation}
such that 
\begin{equation}
   \f{1}{f} \f{\rmd f}{\rmd t} = \gamma.
\end{equation}

But now, suppose that $B(t)$ is a varying function with time, then
\begin{equation}
    \f{1}{f} \f{\rmd f}{\rmd t} = \gamma(t).
\end{equation}
We integrate the equation
\begin{equation}
    \int^{f(t_2)}_{f(t_1)} \f{\rmd f'}{f'}
    = \int^{t_2}_{t_1} \gamma(t) ~ \rmd t,
\end{equation}
and obtain
\begin{equation}
    \ln{\left( \f{f(t_2)}{f(t_1)} \right)}
    = \int^{t_2}_{t_1} \gamma(t) ~ \rmd t.
\end{equation}
Hence
\begin{equation}
    f(t_2) = f(t_1) \exp{\left[ \int^{t_2}_{t_1} \gamma(t) ~ \rmd t \right]}
    = f(t_1) \exp{\left[ \gamma_\text{eff} (t_2 - t_1) \right]},
\end{equation}
and the effective growth rate over the time period is
\begin{equation}
    \gamma_\text{eff} = \f{\int^{t_2}_{t_1} \gamma(t) ~ \rmd t}{t_2 - t_1}.
\end{equation}
The numerator of $\gamma_\text{eff}$ can be found through numerical integration.

\bsp	
\label{lastpage}
\end{document}